\documentstyle[12pt,epsfig]{article}
\def\mathrm{\rm}

\advance\textheight by \topskip
\begin{document}
\tolerance=100000
\thispagestyle{empty}
\setcounter{page}{0}

\newcommand{\be}{\begin{equation}}
\newcommand{\ene}{\end{equation}}
\newcommand{\br}{\begin{eqnarray}}
\newcommand{\er}{\end{eqnarray}}
\newcommand{\ba}{\begin{array}}
\newcommand{\ea}{\end{array}}
\newcommand{\bi}{\begin{itemize}}
\newcommand{\ei}{\end{itemize}}
\newcommand{\bn}{\begin{enumerate}}
\newcommand{\en}{\end{enumerate}}
\newcommand{\bc}{\begin{center}}
\newcommand{\ec}{\end{center}}
\newcommand{\ul}{\underline}
\newcommand{\ol}{\overline}
\newcommand{\ar}{\rightarrow}
\newcommand{\sm}{${\cal {SM}}$}
\newcommand{\mssm}{${\cal {MSSM}}$}
\newcommand{\Dir}{\kern -6.4pt\Big{/}}
\newcommand{\Dirin}{\kern -10.4pt\Big{/}\kern 4.4pt}
\newcommand{\DDir}{\kern -7.6pt\Big{/}}
\newcommand{\DGir}{\kern -6.0pt\Big{/}}

\def\Ord{\buildrel{\scriptscriptstyle <}\over{\scriptscriptstyle\sim}}
\def\OOrd{\buildrel{\scriptscriptstyle >}\over{\scriptscriptstyle\sim}}
\def\pl #1 #2 #3 {{\it Phys.~Lett.} {\bf#1} (#2) #3}
\def\np #1 #2 #3 {{\it Nucl.~Phys.} {\bf#1} (#2) #3}
\def\zp #1 #2 #3 {{\it Z.~Phys.} {\bf#1} (#2) #3}
\def\pr #1 #2 #3 {{\it Phys.~Rev.} {\bf#1} (#2) #3}
\def\prep #1 #2 #3 {{\it Phys.~Rep.} {\bf#1} (#2) #3}
\def\prl #1 #2 #3 {{\it Phys.~Rev.~Lett.} {\bf#1} (#2) #3}
\def\mpl #1 #2 #3 {{\it Mod.~Phys.~Lett.} {\bf#1} (#2) #3}
\def\rmp #1 #2 #3 {{\it Rev.~Mod.~Phys.} {\bf#1} (#2) #3}
\def\ijmp #1 #2 #3 {{\it Int.~J.~Mod.~Phys.} {\bf#1} (#2) #3}
\def\sjnp #1 #2 #3 {{\it Sov.~J.~Nucl.~Phys.} {\bf#1} (#2) #3}
\def\xx #1 #2 #3 {{\bf#1}, (#2) #3}
\def\preprint{{\it preprint}}

\def\brlo{{1}}
\def\brhi{{2}}
\def\gammatot{{3}}
\def\feyndiags{{4}}
\def\sigtot{{5}}
\def\sigtop{{6}}
\def\sigpartons{{7}}
\def\gluons{{8}}
\def\scaledep{{9}}
\def\neventgg{{10}}
\def\neventwh{{11}}
\def\neventwhbb{{12}}
\def\neventllll{{13}}
\def\neventlnln{{14}}

\begin{flushright}
{\large DFTT 34/95}\\ 
{\large DTP/96/100}\\
{\large Cavendish-HEP-96/20}\\
{\large ETH-TH-96/48}\\ 
{\rm November 1996\hspace*{.5 truecm}}\\
\end{flushright}


\begin{center}
{\Large \bf 
Higgs Production at the  LHC: an Update\\ [2mm]
on Cross Sections and Branching
Ratios\footnote{Work supported 
in part by Ministero 
dell' Universit\`a e della Ricerca Scientifica.\\[2. mm]
E-mails:
 Kunszt@itp.phys.ethz.ch;
Moretti@hep.phy.cam.ac.uk; W.J.Stirling@durham.ac.uk.}}\\[0.5cm]
{\large 
Z.~Kunszt$^{a}$, S.~Moretti$^{b,c}$ and W.~J.~Stirling$^{d,e}$}\\[0.15 cm]
{\it a) Theoretical Physics, ETH,}\\
{\it Z\"urich, Switzerland.}\\[0.15cm]
{\it b) Dipartimento di Fisica Teorica, Universit\`a di Torino,}\\
{\it and I.N.F.N., Sezione di Torino,}\\
{\it Via Pietro Giuria 1, 10125 Torino, Italy.}\\[0.15cm]
{\it c) Cavendish Laboratory, University of Cambridge,}\\ 
{\it Madingley Road, Cambridge, CB3 0HE, United Kingdom.}\\[0.15 cm]
{\it d) Department of Physics, University of Durham,}\\
{\it South Road, Durham DH1 3LE, United Kingdom.}\\[0.15cm]
{\it e) Department of Mathematical Sciences, University of Durham,}\\
{\it South Road, Durham DH1 3LE, United Kingdom.}\\[0.15cm]
\end{center}
\begin{abstract}
{\noindent\small 
New theoretical and experimental information motivates a
re-examination of the Standard Model Higgs production rates at the LHC
$pp$ collider.
We present calculations of the relevant cross sections and branching ratios,
including recently calculated QCD next-to-leading order corrections,
new parton distributions fitted to recent HERA structure function data,
and new values for electroweak input parameters, in particular 
for the top quark mass. Cross sections are calculated at two collider
energies, $\sqrt{s} = 10$~TeV and $14$~TeV.}
\end{abstract}
\newpage

\section{Introduction}
\label{sec:intro}

The discovery of the Standard Model (\sm ) Higgs boson is one of the most
important physics goals of the Large Hadron Collider (LHC).
An important prerequisite of LHC Higgs phenomenology is a precise
knowledge of the various production cross sections and decay branching
ratios. Detailed studies (see, for example, Refs.~\cite{guide,LHC})
 have shown that there is no single
production mechanism or decay channel which dominates the phenomenology
over the whole of the relevant Higgs mass range, $O(100\ {\mathrm{GeV}}) <
M_H < O(1\ {\mathrm{TeV}})$, rather there are several different scenarios
depending on the value of $M_H$.

The precision with which such calculations can be performed has improved
significantly over the years. In particular,
\begin{itemize}
\item[{(i)}] next-to-leading order corrections are now known for most
of the subprocess production cross sections and partial decay widths;
\item[{(ii)}]  knowledge of the parton distribution functions has
improved as more precision deep inelastic and other data have become
available;
\item[{(iii)}] the range of possible input parameter values (in particular the
top quark mass $m_t$) has decreased as a result of precision
measurements from LEP, the Tevatron $p \bar p$ collider and other
experiments.
\end{itemize}

\noindent
As a consequence, many of the numerical results to be found in the literature
are now out-of-date. We are therefore motivated to update
the calculations  \cite{oldpaper}
of the relevant cross sections and branching ratios to
take into account the improvements discussed above.
The output of our analysis will be a set of benchmark results for cross
sections and event rates as a function of $M_H$, for the two
`standard' LHC collision energies, $\sqrt{s} = 10$ and $14\ {\mathrm{TeV}}$.
Note that we are not attempting to perform a detailed analysis of
signals, backgrounds and search strategies. In this respect, by far the
most complete studies to date can be found in the recent ATLAS
\cite{ATLAS} and CMS \cite{CMS}
Technical Proposals. Even there, however, one finds slight
inconsistencies in the way the cross sections and branching ratios are
calculated (leading versus next-to-leading order cross sections,
out-of-date parton distributions and parameter values, etc.).
The present study will enable the Higgs production
cross sections, event rates and significance factors
used in Refs.~\cite{ATLAS} and \cite{CMS} to be renormalised to the
most up-to-date values.

Another important factor is the theoretical uncertainty of the
predictions. In most cases we can estimate these by varying appropriate
 input quantities like the parton distributions, parameter values and
renormalisation and factorisation scales. As a result, we can
identify those quantities where more theoretical work is needed to
improve the precision.

The paper is organised as follows. In the following Section we list
and discuss the set of QCD and electroweak input parameters for the
calculations. In Section \ref{sec:br} we calculate the complete set of
branching ratios needed to predict event rates for specific channels.
Section~\ref{sec:cross} contains the main results of the paper:
numerical cross section calculations for a variety of Higgs production
and decay processes. Finally, our conclusions and outlook are presented
in Section~\ref{sec:conc}.

\section{Electroweak and QCD input parameters}
\label{sec:input}

\noindent
Most of the discrepancies in  the literature concerning the values 
of Higgs cross sections and branching ratios arises simply from 
different choices 
of the electroweak and QCD  input parameters. For reference, therefore,
 we list here the numerical values adopted in this study:
$$M_Z=91.186~{\mathrm {GeV}},\quad\quad \Gamma_Z=2.495~{\mathrm {GeV}},$$  
$$M_W=80.356~{\mathrm {GeV}},\quad\quad \Gamma_W=2.088~{\mathrm {GeV}},$$ 
\be\label{ewparam}
G_F=1.16639\times10^{-5}~{\mathrm {GeV}}^{-2},
\quad\quad\alpha_{em}\equiv \alpha_{em}(M_Z)= 1/128.9 .
\ene
Although each of these parameters has a small measurement error, 
the effect of these on the event rates computed below is negligible
compared to other uncertainties.
The charged and neutral weak fermion--boson couplings are defined by
\be\label{GF}
g_W^2 = \frac{e^2}{\sin^2\theta_W} =  4 \sqrt{2} G_F M_W^{2}, \qquad
g_Z^2 = \frac{e^2}{\sin^2\theta_W\; \cos^2\theta_W} =  4 \sqrt{2} G_F M_Z^{2}. \qquad
\ene
For the vector and axial couplings
of the $Z$ boson to fermions, we use the `effective leptonic' value
\be\label{s2w_eff}
\sin^2_{\rm {eff}}(\theta_W)=0.232.
\ene
The QCD strong coupling enters explicitly in the
 production cross sections and in the branching ratios,
 and implicitly in the parton distributions. 
Since most quantities we calculate are known to next-to-leading order,
unless otherwise stated we  use    $\alpha_s$ evaluated
at two-loop order, with\footnote{The 
values of $\Lambda^{(n_f)}_{\overline{{MS}}}$
for other  $n_f$'s are calculated 
according to the prescription in 
Ref.~\cite{MARCIANO}.} $\Lambda^{(4)}_{\overline{{MS}}}=230$~MeV  
to match  our default parton distribution set,
 and with a scale $\mu$ chosen appropriately for the process in question.
 The choice of scale for production cross sections is discussed in
 Section~4 below,  while for the branching 
ratios we adopt  the prescriptions of
Refs.~\cite{SMHqqQCDcorr,SMHVphQCDcorr,SMHggQCDcorr}.
Our default  parton distributions are the MRS(A) set
\cite{MRSA}, which have been fitted to a wide range of HERA
and other deep inelastic scattering 
data. We also display results for the recent MRS(R1,R2) parton sets 
\cite{mrs96fit}.
They represent an update of the MRS(A) set and are 
fitted to the latest HERA data. They also enable
us to study the dependence of the production cross sections on the value
of $\alpha_s$ in a consistent manner.\footnote{When we  compare the cross
sections obtained by using different sets of parton distributions, we 
of course use the appropriate $\Lambda^{(n_f)}_{\overline{{MS}}}$
values.}

For the fermion masses we  take
$m_\mu=0.105$~GeV, $m_\tau=1.78$~GeV, $m_s=0.3$~GeV, $m_c = 1.4$~GeV,
 $m_b=4.25$~GeV
and $m_t=175$~GeV, with all decay widths equal to 
zero except
for $\Gamma_t$. We calculate this at tree-level
within the \sm, using the expressions
given in Ref.~\cite{widthtopSM}. We study the variation of the production
cross sections with $m_t$ in the range $165 < m_t < 185$~GeV, which subsumes
the recent direct measurement (CDF and D0 combined)
from the Tevatron $p \bar p$ collider
of $m_t = 175 \pm 6$~GeV \cite{topmass96}.
The first generation of fermions and all neutrinos are taken to be
massless, i.e. $m_u=m_d=m_e=m_{\nu_e}=0$ and
$m_{\nu_\mu}=m_{\nu_\tau}=0$, with zero decay widths.

We assume that $M_H =  80$~GeV is a conservative
 discovery mass limit for LEP2, and therefore focus our attention
on the mass range $80\ {\mathrm{GeV}} \leq M_H \leq 1\ {\mathrm{TeV}}$.
The discussion naturally falls into classes, depending on whether
$M_H$ is less than or greater than $O(2 M_W)$\footnote{This defines
the so-called `intermediate-mass' and `heavy' Higgs mass ranges,
respectively.}.

\section{\sm\ Higgs branching ratios}
\label{sec:br}

\noindent
The branching ratios of the \sm\ Higgs boson have been 
studied in many papers. 
A useful compilation of the early works on this subject can be found in
Ref.~\cite{guide}, where all the most relevant formulae
for on-shell decays are summarised.
Higher-order corrections
to  most  of the decay processes have also been computed
(for
up-to-date reviews  see Refs.~\cite{corrreview,lep2review} 
and references therein), 
as well as 
the rates for the off-shell decays $H\ar W^{*}W^{*},Z^{*}Z^{*}$ 
\cite{SMHVVoff}, $H\ar Z^{*}\gamma$ \cite{SMHVphoff} and $H\ar t^*\bar t^*$  
\cite{SMHtt_HVVoff}. Threshold effects due to the possible
formation of $t\bar t$ bound states in the one-loop induced process
$H\ar \gamma\gamma$ have also been studied \cite{threshold}. 

In our calculations we include  only the (large) 
QCD corrections to the  \sm\ Higgs partial widths into heavy quark 
pairs \cite{SMHqqQCDcorr} and
into $Z\gamma$, $\gamma\gamma$ \cite{SMHVphQCDcorr} and $gg$ 
\cite{SMHggQCDcorr}\footnote{The higher-order electroweak corrections and
their interplay with the QCD corrections 
do not significantly change the branching ratios which
are phenomenologically relevant for Higgs searches at the LHC.}. Since the
QCD corrections to the top loops in 
$Z\gamma$, $\gamma\gamma$ and $gg$ decays, as 
given in Ref.~\cite{SMHVphQCDcorr},
 are
valid only for $M_{H}\ll 2m_t$, we have implemented these only far below
the $t\bar t$ threshold, which is in fact  the only region where
these decay channels could be important\footnote{Numerical results
valid for any value of the ratio $\tau=M_H^2/4m_t^2$
have recently been presented \cite{allspectrum}.}.

The bulk of the QCD corrections to $H\ar q\bar q$ can 
be absorbed into a `running' quark mass $m_q(\mu)$, evaluated at the energy 
scale $\mu=M_H$ (for example). 
The importance of this effect for the case $q=b$, 
with respect to intermediate-mass Higgs searches at the LHC, 
has been discussed in Ref.~\cite{oldpaper}.
There is, however, a  slight subtlety concerning 
$t \bar t$ decays \cite{MSSMpaper}.
For  $H \ar q \bar q$ decays involving light quarks 
($q=s,c,b$), the use of the running
quark mass $m_q(\mu = M_H)$ takes into account large logarithmic
corrections at higher orders in QCD perturbation theory
\cite{SMHqqQCDcorr},  and so in principle one could imagine using the
same procedure for  $H \ar t \bar t$, at least in the limit $M_H
\gg m_t$. In practice, however, we are interested only in the
region $M_H/m_t  \sim  O(1)$. In the case of the top quark loop mediated
decay $H \ar gg$, it is well known that the higher-order QCD corrections
are minimised if the quark mass is defined at the pole of the
propagator, i.e. $m_t(\mu = m_t)$  \cite{running}. 
To be  consistent, therefore, we use 
the {\it same} top mass $m_t(\mu=m_t)$ in the decay width for $H
\ar t \bar t$ (and $H\ar Z\gamma,\gamma\gamma$ as well). For the light quark 
loop contributions to the
 $H\ar gg(Z\gamma,\gamma\gamma)$ decay widths
 we use pole(running) masses (defined at the scale $\mu=M_H$)
 \cite{running,emailSpira}.

Our results on the Higgs branching ratios are summarised in 
Figs.~\brlo--\gammatot.
Fig.~\brlo\ shows the branching ratios, for $M_H \leq 200$~GeV,
for the channels: (a) $b\bar b$,
$c\bar c$, $\tau^+\tau^-$, $\mu^+\mu^-$ and $gg$; and (b) $WW$,
$ZZ$, $\gamma\gamma$ and $Z\gamma$. 
The patterns of the various curves are not significantly different
from those presented in Ref.~\cite{oldpaper}. The
inclusion of the QCD corrections in the quark-loop induced
decays (which apart from small changes in the parameter values
is the only
significant difference with respect to the calculation in \cite{oldpaper})
turns out to give a variation of at most a few per cent for the
decays $H\ar\gamma\gamma$
and $H\ar Z\gamma$, while for $H\ar gg$ differences are of order
50--60\%. However this has little phenomenological relevance,
since this decay width makes a negligible contribution to the total width,
and is an unobservable channel in practice.

Note that for the below-threshold
decays $H\ar W^*W^*$ and $H\ar Z^*Z^*$, we do not constrain
 one of the vector bosons to be
on-shell when the decaying Higgs boson mass exceeds the gauge
boson rest mass. Instead we integrate numerically over 
the virtualities of both
decay products, see for example Ref.~\cite{MSSMpaper},
thus avoiding errors in the threshold region.
  
Fig.~\brhi\ shows the \sm\ Higgs branching ratios for
200~GeV $\le M_H \le$ 1000~GeV. Apart from our  use  of a different
$m_t$ value, the curves show the same
distinctive features as those presented in Ref.~\cite{oldpaper}.
The inclusion  here of below-threshold  
$H\ar t^*\bar t^*$ decays does not give any observable
effect, since this channel  is heavily suppressed by the $WW$
and $ZZ$ decays.

For completeness, we show in Fig.~\gammatot\ the total \sm\ Higgs decay width 
over the range 50~GeV $\leq M_H \leq$ 1000~GeV.
For further comments on the implications of the various \sm\ Higgs branching
ratios on the various search strategies, we refer the reader to 
Ref.~\cite{oldpaper}.

\section{\sm\ Higgs production cross sections and event rates}
\label{sec:cross}
 
There are only a few Higgs production mechanisms which lead to detectable 
cross sections at the LHC. Each of them makes use of the preference of
the \sm\ Higgs to couple to heavy particles: either massive vector
bosons ($W$ and $Z$) or massive quarks (especially $t$-quarks).
They are (see Fig.~\feyndiags):
\begin{itemize}
\item[{(a)}]  gluon-gluon fusion \cite{xggh}, 
\item[{(b)}]  $WW$, $ZZ$ fusion \cite{xvvh},
\item[{(c)}]  associated production with $W$, $Z$ bosons \cite{xvh},
\item[{(d)}]  associated production with $t\bar t$ pairs \cite{xtth}.
\end{itemize}
A complete review on the early literature on $pp$ collider
 \sm\ Higgs boson phenomenology, based on these production mechanisms, 
 can be found in  Ref.~\cite{guide}. 
 
There are various uncertainties in the rates of the above 
processes, although none is particularly large. The most significant
are: (i) the lack of precise knowledge  of the gluon
distribution at small $x$, which is important
for the intermediate-mass Higgs, and (ii)  the effect of
unknown  higher-order perturbative QCD corrections.
In what follows, we will attempt to quantify the former by using 
recent
sets of different
parton distributions \cite{MRSA,mrs96fit,CTEQ3,GRV94,MRSG,MRSalphas} which give
excellent fits to a wide range of deep inelastic scattering data 
(including the new structure function data from the  HERA $ep$ collider)
and to data on other hard scattering processes. The latter will be
estimated by studying the dependence (at next-to-leading
order) of the results 
 on the values of the renormalisation and factorisation scales.

The next-to-leading order  QCD corrections  are known
for processes (a), (b) and (c) and are included
in our calculations. By far the most important of these
are the corrections to the gluon fusion process (a) which have
been calculated in Ref.~\cite{newKfacgg}.
In the limit  where  the Higgs mass is far below the
$2m_t$ threshold, these corrections are calculable analytically
\cite{Kfacgg1,Kfacgg2,Kfacgg3}.
In fact, it turns  out that the analytic result is a good approximation
over the complete $M_H$ range, and so we will use it in our analysis
\cite{DjouadiRev,sdgz}.
In Ref.~\cite{sdgz} the impact of the next-to-leading order QCD corrections
for the gluon fusion process on LHC cross sections was investigated,
both for the \sm\ and for the \mssm. Where our calculations overlap,
we find agreement with the results of \cite{sdgz}.

Overall, the next-to-leading order correction increases 
the leading-order result\footnote{Unless otherwise stated,
here and in what follows we use 
`leading-order' to denote the cross 
section calculated using the leading-order matrix element
combined with next-to-leading order $\alpha_s$  and
parton distributions.} by a factor of about 2, when the normalisation
and factorisation scales are set equal to our default choice $\mu = M_H$.
This `$K$-factor' can be traced to a large constant piece in the
next-to-leading correction \cite{approxKfac},
\be
\label{Kfac}
K \approx 1 + {\alpha_s(\mu=M_H)\over \pi}\;\left[{\pi^2}+{11\over
2}\right] .
\ene
Such a large  $K$-factor usually implies a non-negligible scale
dependence of the theoretical cross section. We will investigate this further 
below.

The next-to-leading order corrections to
 the  $VV$ fusion \cite{KfacVV} and $VH$ \cite{KfacVH} 
production cross sections ($V=W,Z$) are quite small, increasing the total
cross sections by no more than $\approx10\%$ (at large $M_H$)
 and $\approx20\%$, respectively. 
Note that for the former we follow Ref.~\cite{KfacVV} and choose the
factorisation scale to be $\mu^2 = -q_V^2$, where $q^\mu_V$ is the 
four-momentum of the virtual $V=W,Z$ boson. For the latter we choose
$\mu^2 = {\mathaccent 94{s}_{pp}} \approx (p_V+p_H)^2$ --- the scale
dependence here is in fact very weak.
The  corresponding QCD corrections for  the $t\bar tH$ mechanism have not 
yet been computed and so we use the leading-order matrix elements with
$\mu^2 = {\mathaccent 94{s}_{pp}}$.
 A recent review of the higher-order corrections to Higgs
cross sections  can be found in Ref.~\cite{DjouadiRev}. 

Our results for the production cross sections are given in
Figs.~\sigtot(a) and (b), for LHC energies of 10 and 14~TeV respectively.
These figures can be directly compared to Fig.~6 of \cite{oldpaper}, where
the cross sections were evaluated at 16~TeV.
The pattern of the various curves is largely unchanged, the main differences
coming from the updated input parameters and parton distributions.
The gluon-gluon fusion
mechanism is   dominant over all
the Higgs mass range, followed by $WW$/$ZZ$ fusion which
becomes comparable in magnitude to  gluon-gluon fusion for
very large Higgs masses.
The cross sections of the other production mechanisms ($WH$, $ZH$ and 
$t\bar t H$) are much smaller, by  between one
($M_H \sim 50$~GeV) and almost three ($M_H \sim 1000$~GeV)
orders of magnitude.

The values of all the cross sections  at 
$\sqrt s_{pp}=10$ and 14~TeV have of course changed with respect to
their values at 16~TeV. The decrease in the cross sections between
16~TeV and 14~TeV is quite small, with a more substantial 
decrease between 14~TeV and 10~TeV. We quantify this in Fig.~\sigtot(c),
which shows the ratio of cross sections $\sigma(10\; {\mathrm{TeV}})/\sigma(14
\; {\mathrm{TeV}})$.
The effect is largely due to the decrease in the incoming
parton luminosities as higher values of momentum fraction $x$ are probed
as the overall collider energy is smaller. The effect is therefore more marked
for gluon-induced processes (compare the reductions for $gg\to H$
and $q \bar q \to WH$), and for processes with a higher subprocess
energy at the same Higgs mass  (compare the reductions for $gg\to H$
and $gg\to t \bar t H$).
For reference, the numerical values of the cross sections displayed
in Figs.~\sigtot(a) and (b) are listed in Tables~1 and 2 respectively. 

\begin{table}
\begin{center}
\begin{tabular}{|c|c|c|c|c|c|}
\hline
\rule[0cm]{0cm}{0cm}
$M_H$ (GeV)                           &\omit  
$~$                                   &\omit  
$~$                                   &\omit  
$\sigma(pp\rightarrow H X)$ (pb)      &\omit  
$~$                                   &
$~$                                   \\ \hline  \hline
\rule[0cm]{0cm}{0cm}
$~$                                           &  
$gg\rightarrow H$                             &  
$gg,q\bar q\rightarrow t\bar tH$              &  
$qq\rightarrow qqH$                           &  
$q\bar q'\rightarrow WH$                      & 
$q\bar q \rightarrow ZH$              \\ \hline\hline
\rule[0cm]{0cm}{0cm}
$50$ &  
$88.$  &    
$1.1$  &  
$4.7$  &  
$13.$  &  
$6.1$  \\ \hline
\rule[0cm]{0cm}{0cm}
$100$ &  
$27.$  &  
$0.26$  &    
$2.9$  &  
$2.0$  &  
$1.0$  \\ \hline
\rule[0cm]{0cm}{0cm}
$150$ &  
$13.$  &  
$8.2\times10^{-2}$  &    
$1.9$  &  
$0.55$  &  
$0.29$  \\ \hline
\rule[0cm]{0cm}{0cm}
$200$ &  
$7.4$  &  
$3.3\times10^{-2}$  &    
$1.3$  &  
$0.20$  &  
$0.11$  \\ \hline
\rule[0cm]{0cm}{0cm}
$250$ &  
$5.0$  &  
$1.6\times10^{-2}$  &    
$0.97$  &  
$8.7\times10^{-2}$  &  
$4.7\times10^{-2}$  \\ \hline
\rule[0cm]{0cm}{0cm}
$300$ &  
$3.8$  &  
$9.2\times10^{-3}$  &    
$0.71$  &  
$4.3\times10^{-2}$  &  
$2.3\times10^{-2}$  \\ \hline
\rule[0cm]{0cm}{0cm}
$350$ &  
$3.7$  &  
$5.9\times10^{-3}$  &    
$0.54$  &  
$2.3\times10^{-2}$  &  
$1.3\times10^{-2}$  \\ \hline
\rule[0cm]{0cm}{0cm}
$400$ &  
$3.7$  &  
$4.2\times10^{-3}$  &    
$0.42$  &  
$1.3\times10^{-2}$  &  
$7.2\times10^{-3}$  \\ \hline
\rule[0cm]{0cm}{0cm}
$450$ &  
$2.7$  &  
$3.0\times10^{-3}$  &    
$0.32$  &  
$8.2\times10^{-3}$  &  
$4.3\times10^{-3}$  \\ \hline
\rule[0cm]{0cm}{0cm}
$500$ &  
$1.8$  &  
$2.2\times10^{-3}$  &    
$0.26$  &  
$5.2\times10^{-3}$  &  
$2.7\times10^{-3}$  \\ \hline
\rule[0cm]{0cm}{0cm}
$550$ &  
$1.2$  &  
$1.7\times10^{-3}$  &    
$0.21$  &  
$3.4\times10^{-3}$  &  
$1.8\times10^{-3}$  \\ \hline
\rule[0cm]{0cm}{0cm}
$600$ &  
$0.79$  &  
$1.3\times10^{-3}$  &    
$0.17$  &  
$2.3\times10^{-3}$  &  
$1.2\times10^{-3}$  \\ \hline
\rule[0cm]{0cm}{0cm}
$650$ &  
$0.53$  &  
$1.0\times10^{-3}$  &    
$0.13$  &  
$1.6\times10^{-3}$  &  
$8.1\times10^{-4}$  \\ \hline
\rule[0cm]{0cm}{0cm}
$700$ &  
$0.36$  &  
$7.7\times10^{-4}$  &    
$0.11$  &  
$1.1\times10^{-3}$  &  
$5.6\times10^{-4}$  \\ \hline
\rule[0cm]{0cm}{0cm}
$750$ &  
$0.25$  &  
$6.1\times10^{-4}$  &    
$9.1\times10^{-2}$  &  
$7.9\times10^{-4}$  &  
$3.9\times10^{-4}$  \\ \hline
\rule[0cm]{0cm}{0cm}
$800$ &  
$0.17$  &  
$4.7\times10^{-4}$  &    
$7.5\times10^{-2}$  &  
$5.7\times10^{-4}$  &  
$2.8\times10^{-4}$  \\ \hline
\rule[0cm]{0cm}{0cm}
$850$ &  
$0.12$  &  
$3.6\times10^{-4}$  &    
$6.3\times10^{-2}$  &  
$4.2\times10^{-4}$  &  
$2.1\times10^{-4}$  \\ \hline
\rule[0cm]{0cm}{0cm}
$900$ &  
$8.5\times10^{-2}$  &  
$2.9\times10^{-4}$  &    
$5.2\times10^{-2}$  &  
$3.1\times10^{-4}$  &  
$1.5\times10^{-4}$  \\ \hline
\rule[0cm]{0cm}{0cm}
$950$ &  
$6.1\times10^{-2}$  &  
$2.3\times10^{-4}$  &  
$4.4\times10^{-2}$  &  
$2.3\times10^{-4}$  &  
$1.1\times10^{-4}$  \\ \hline
\rule[0cm]{0cm}{0cm}
$1000$ &  
$4.4\times10^{-2}$  &  
$1.9\times10^{-4}$  &  
$3.7\times10^{-2}$  &  
$1.8\times10^{-4}$  &  
$8.3\times10^{-5}$  \\ \hline
\end{tabular}
\caption{Higgs production cross sections for $pp$ collisions
at $\protect\sqrt s_{pp} = 10$~TeV.}
\end{center}
\end{table}


\begin{table}
\begin{center}
\begin{tabular}{|c|c|c|c|c|c|}
\hline
\rule[0cm]{0cm}{0cm}
$M_H$ (GeV)                           &\omit  
$~$                                   &\omit  
$~$                                   &\omit  
$\sigma(pp\rightarrow H X)$ (pb)      &\omit  
$~$                                   &
$~$                                   \\ \hline  \hline
\rule[0cm]{0cm}{0cm}
$~$                                           &  
$gg\rightarrow H$                             &  
$gg,q\bar q\rightarrow t\bar tH$              &  
$qq\rightarrow qqH$                           &  
$q\bar q'\rightarrow WH$                      & 
$q\bar q \rightarrow ZH$              \\ \hline\hline
\rule[0cm]{0cm}{0cm}
$50$ &  
$139.$  &  
$2.6$  &  
$7.8$  &  
$19.$  &  
$9.3$  \\ \hline
\rule[0cm]{0cm}{0cm}
$100$ &  
$45.$  &  
$0.60$  &  
$5.1$  &  
$3.1$  &  
$1.6$  \\ \hline
\rule[0cm]{0cm}{0cm}
$150$ &  
$23.$  &  
$0.20$  &  
$3.5$  &  
$0.86$  &  
$0.47$  \\ \hline
\rule[0cm]{0cm}{0cm}
$200$ &  
$14.$  &  
$8.2\times10^{-2}$  &  
$2.5$  &  
$0.32$  &  
$0.18$  \\ \hline
\rule[0cm]{0cm}{0cm}
$250$ &  
$9.6$  &  
$4.3\times10^{-2}$  &  
$1.8$  &  
$0.14$  &  
$8.0\times10^{-2}$  \\ \hline
\rule[0cm]{0cm}{0cm}
$300$ &  
$7.6$  &  
$2.6\times10^{-2}$  &  
$1.4$  &  
$7.4\times10^{-2}$  &  
$4.1\times10^{-2}$  \\ \hline
\rule[0cm]{0cm}{0cm}
$350$ &  
$7.5$  &  
$1.8\times10^{-2}$  &  
$1.1$  &  
$4.1\times10^{-2}$  &  
$2.2\times10^{-2}$  \\ \hline
\rule[0cm]{0cm}{0cm}
$400$ &  
$7.8$  &  
$1.3\times10^{-2}$  &  
$0.87$  &  
$2.4\times10^{-2}$  &  
$1.3\times10^{-2}$  \\ \hline
\rule[0cm]{0cm}{0cm}
$450$ &  
$5.8$  &  
$1.0\times10^{-2}$  &  
$0.70$  &  
$1.5\times10^{-2}$  &  
$8.2\times10^{-3}$  \\ \hline
\rule[0cm]{0cm}{0cm}
$500$ &  
$4.1$  &  
$7.8\times10^{-3}$  &  
$0.57$  &  
$9.9\times10^{-3}$  &  
$5.3\times10^{-3}$  \\ \hline
\rule[0cm]{0cm}{0cm}
$550$ &  
$2.8$  &  
$6.3\times10^{-3}$  &  
$0.47$  &  
$6.7\times10^{-3}$  &  
$3.6\times10^{-3}$  \\ \hline
\rule[0cm]{0cm}{0cm}
$600$ &  
$1.9$  &  
$5.1\times10^{-3}$  &  
$0.39$  &  
$4.7\times10^{-3}$  &  
$2.5\times10^{-3}$  \\ \hline
\rule[0cm]{0cm}{0cm}
$650$ &  
$1.3$  &  
$4.0\times10^{-3}$  &
$0.33$  &  
$3.3\times10^{-3}$  &  
$1.8\times10^{-3}$  \\ \hline
\rule[0cm]{0cm}{0cm}
$700$ &  
$0.91$  &  
$3.2\times10^{-3}$  &  
$0.27$  &  
$2.4\times10^{-3}$  &  
$1.3\times10^{-3}$  \\ \hline
\rule[0cm]{0cm}{0cm}
$750$ &  
$0.64$  &  
$2.6\times10^{-3}$  &  
$0.23$  &  
$1.8\times10^{-3}$  &  
$9.2\times10^{-4}$  \\ \hline
\rule[0cm]{0cm}{0cm}
$800$ &  
$0.46$  &  
$2.1\times10^{-3}$  &  
$0.20$  &  
$1.3\times10^{-3}$  &  
$6.9\times10^{-4}$  \\ \hline
\rule[0cm]{0cm}{0cm}
$850$ &  
$0.33$  &  
$1.7\times10^{-3}$  &  
$0.17$  &  
$1.0\times10^{-3}$  &  
$5.2\times10^{-4}$  \\ \hline
\rule[0cm]{0cm}{0cm}
$900$ &  
$0.24$  &  
$1.4\times10^{-3}$  &  
$0.15$  &  
$7.8\times10^{-4}$  &  
$3.9\times10^{-4}$  \\ \hline
\rule[0cm]{0cm}{0cm}
$950$ &  
$0.18$  &  
$1.2\times10^{-3}$  &  
$0.13$  &  
$6.0\times10^{-4}$  &  
$3.0\times10^{-4}$  \\ \hline
\rule[0cm]{0cm}{0cm}
$1000$ &  
$0.13$  &  
$9.6\times10^{-4}$  &  
$0.11$  &  
$4.7\times10^{-4}$  &  
$2.3\times10^{-4}$  \\ \hline
\end{tabular}
\caption{Higgs production cross sections for $pp$ collisions
at $\protect\sqrt s_{pp} = 14$~TeV.}
\end{center}
\end{table}

In early phenomenological studies, such as those of 
 Refs.~\cite{guide,oldpaper}, the uncertainty in the mass of the 
 (as yet undiscovered) top quark provided a significant 
 additional theoretical uncertainty in the $gg\to H$ and 
 $gg,q\bar q\rightarrow t\bar tH$ cross sections. However, this
 source of uncertainty has largely disappeared: the most recent
 experimental measurement is $m_t = 175 \pm 6$~GeV \cite{topmass96},
 and the corresponding uncertainty on the Higgs cross sections is quite
 small. This is illustrated in 
 Fig.~\sigtop, which   shows  the 
cross section ratios $\sigma(m_t = 165\; {\mathrm {GeV}})
/\sigma(m_t = 175\; {\mathrm {GeV}})$ and 
 $\sigma(m_t = 185\; {\mathrm {GeV}})
/\sigma(m_t = 175\; {\mathrm {GeV}})$
 for $gg\ar H$ and $gg, q\bar q\ar t\bar t H$
 at $\sqrt s_{pp}=14$~TeV. The curves have the same
 qualitative features as those (for $m_t = 150, 175, 200$~GeV) in  
Fig.~6 of Ref.~\cite{oldpaper}, although the spread is of course much smaller.
 A lighter top  quark gives a bigger 
cross section for $gg, q\bar q \to t \bar t H$ over essentially all of
the  Higgs
mass range and for $gg\to H$ for $M_H \Ord 400$~GeV. A heavier top
enhances the $gg \to H$ cross section at large $M_H$, where the 
increase of the $t \bar t H$ coupling is the dominant factor.
For  very heavy Higgs bosons, the cross section increases by 
about 30\% as $m_t$ increases from 165 to 185~GeV. If we assume that by the time
the LHC comes into operation the top quark mass will be known with
a precision of $\pm 5$~GeV or better, then the residual uncertainty
in the Higgs production cross section will be less than $\pm O(10\%)$. 

Since a large part of the uncertainties on the Higgs cross sections
at the LHC stems from  the small $x$ behaviour of the gluon
distributions\footnote{For a simple fusion process
like $gg \to H$, the parton $x$ value is typically $x\sim M_H/\sqrt s_{pp}$.},
Fig.~\sigpartons\ shows the total rates for the $gg\ar H$ process
at $\sqrt s_{pp}=14$~TeV
for 11 different  sets of parton 
distributions: MRS(A) \cite{MRSA}, MRS(A$'$),
MRS(G) \cite{MRSG}, MRS(R1,R2) \cite{mrs96fit}, 
CTEQ2M, CTEQ2MS, CTEQ2MF, CTEQ2ML, CTEQ3M 
\cite{CTEQ3} and GRV94HO \cite{GRV94}. Evidently,
a band of $\pm 20\%$ centred on the MRS(A) calculation
 covers the various cross section predictions.
 Fig.~\gluons\ shows the corresponding  gluon distributions for  the same 
 parton sets
used in  Fig.~\sigpartons, in the $x$ range spanned by 
 intermediate- and heavy-mass
 Higgs production, and at the characteristic factorisation 
 scale $Q = x \sqrt s_{pp}$.

There are two main factors which influence  the differences
between the various predictions shown in Fig.~\sigpartons:
the shape of the gluon distribution at medium and small $x$, and
the value of $\alpha_s$ associated with each parton set.
The majority of parton sets in Fig.~\sigpartons\ use the traditional
`DIS' $\alpha_s$ value, i.e.   $\alpha_s(M_Z)\simeq 0.113$.
 For such sets, the differences in the cross sections come mainly
 from the shape of the starting gluon distribution. Thus the sets
 MRS(A,A$'$) and CTEQ3M yield similar predictions, being fitted
 to similar data sets under similar assumptions.
 The gluons of sets MRS(G) and GRV94HO have significantly steeper gluons
 (see Fig.~\gluons),
 and hence give larger Higgs cross sections at small $M_H$.
In fact, both these gluons have recently been shown 
to be in disagreement with the the latest HERA data \cite{mrs96fit}.
Note also that the CTEQ2M$x$ sets of partons give generally small
Higgs cross sections due to smaller gluons at small $x$. These too are
disfavoured by the latest HERA data.

The most recent partons shown in Fig.~\sigpartons\
are the sets MRS(R1) amd MRS(R2). The former are effectively the 
successors to the MRS(A,A$'$) partons, and give similar Higgs cross sections.
For MRS(R1), $\alpha_s(M_Z)=  0.113$ which is slightly larger than the
value for MRS(A,A$'$). In addition, the starting gluon for MRS(R1)is also
slightly larger than that for  MRS(A$'$)
 (see Fig.~5(a) of Ref.~\cite{mrs96fit}). These two effects combine to 
 give a MRS(R1)  Higgs cross section which is between 5\% and 10\% larger
than that of MRS(A). More interesting is the prediction for 
the MRS(R2) partons. For this set $\alpha_s(M_Z)=  0.120$.
This larger value is more in line with the LEP $e^+e^- \to$hadrons/jets
determinations, and also with the CDF and D0 large $E_T$ jet data
(for a full discussion see Ref.~\cite{mrs96fit}). The effect on the Higgs
cross section is very noticeable, see Fig.~\sigpartons. The MRS(R2) 
cross section is some 10$-$15\% larger than that of MRS(R1), consistent
with the difference between the values of $\alpha_s^2$ 
corresponding to each set.

 It should of course be remembered that by the time
 the  LHC comes into operation, the uncertainty on the gluon distribution
at medium and small $x$ may be  expected to be significantly smaller, 
principally
due to improved measurements of the small-$x$  deep-inelastic structure 
functions at HERA, and of large $p_T$ jet and prompt photon production at the
Tevatron $p \bar p $ collider (see for example Refs.~\cite{MRSG,mrs96fit}).
The apparent `disagreement' between the DIS and LEP $\alpha_s$ values
will also presumably be resolved.
At the present time, we may say that the $\pm 20\%$ spread
around the MRS(A) prediction in Fig.~\sigpartons\ 
constitutes a conservative estimate of the uncertainty on
 the Higgs cross section predictions due to 
 parton distributions.\footnote{The parton distribution uncertainty
 on the sub-dominant quark-induced Higgs cross sections,
 e.g. $ q q \to qqH$, is of course much smaller, as the quark distributions
 are pinned down rather precisely by DIS structure function data.}

As already mentioned, the dominant $gg\to H$ process has a large
next-to-leading order correction, which leads to a non-negligible
scale dependence.
Fig.~\scaledep\ shows the dependence at lowest and higher order
of the  Higgs production cross 
section in the gluon-gluon channel
on the (equal) renormalisation and factorisation scales $\mu$, e.g.,
for $\sqrt{s} = 14$~TeV, $M_H = 100$~GeV and $m_t=175$ GeV.
To make a consistent analysis, we have plotted the rates obtained
by using, on the one hand (continuous line), NLO amplitude formulae, 
NLO parton distributions (i.e. GRV94HO) and $\alpha_s$ computed 
at two loops and, on the other hand (dashed line), the LO matrix element, 
LO structure functions (i.e. GRV94LO) and strong coupling 
constant evolved at one loop.
We use the  GRV94 sets of parton distributions in order
to allow for a more straightforward comparison of our results 
with the corresponding ones given in Ref.~\cite{sdgz}, although
one should notice that in Ref.~\cite{sdgz} the old GRV92 set  
\cite{GRV92} was used, 
so that this might be in the end a source of small differences. 
As can be seen from the figure, the (unphysical) variation of the
cross section with the two scales
is largely reduced at higher order, as expected.
 If $\mu$ is varied (conservatively) between $M_H/4$ 
and $4M_H$, the rates at NLO decrease by a factor 1.54, whereas at LO the 
ratio is 1.93. Furthermore, we have repeated the calculations presented
in Ref.~\cite{sdgz},  using $M_H=150$ and 500 GeV (and also, for 
consistency, $m_t=174$ GeV ). Our results exhibit 
the same pattern recognised there. That is,  the  improvement in  
scale stability  gained
at next-to-leading order is more significant for large Higgs masses.
However, we  find that our NLO rates are typically more sensitive to
the value of $\mu$  than those given in Ref.~\cite{sdgz} (although by only a few
percent, in general).
We believe that this difference originates in our use of
 the analytical formulae obtained in the heavy top approximation
$M_H^2/4m_t^2\ll 1$, whereas
in Ref.~\cite{sdgz} the exact (numerical) results were presented.
In fact, we know that the scale dependence at NLO of approximated
results is a delicate issue, since  in several
 instances these have been found to be
more sensitive to the choice of $\mu$ that those at LO (see, for example,
Ref.~\cite{Kfacgg3})\footnote{Indeed we have been able to reproduce the
trend of some of the results given in Ref.~\cite{Kfacgg3}.}.
Note that lowering the collider energy slightly enhances the 
$\mu$-dependence of the cross sections. For example, at 10~TeV the numbers 
corresponding to the two ratios mentioned above are 1.60 (NLO) and 2.08 (LO), 
respectively.

Next, we multiply the production cross sections by the branching
ratios of Section~\ref{sec:br} to obtain event rates for various channels.
Considering all the possible combinations of
production mechanisms and decay channels
\cite{ATLAS,CMS},
the best chance of discovering a \sm\ Higgs at the LHC appears 
to be  given by the 
following signatures: 
(i) $gg\ar H\ar \gamma\gamma$, (ii)
$q\bar q'\ar WH\ar \ell\nu_\ell\gamma\gamma$\footnote{In principle
we should also include  $q\bar q\ar ZH\ar
\ell^+\ell^-\gamma\gamma$, although a very high luminosity would be
needed to make this detectable.} and (iii) $gg\ar H\ar
Z^{(*)}Z^{(*)}\ar \ell^+\ell^-\ell'^{+}\ell'^{-}$, where $\ell,\ell'=e$
or $\mu$.
Recently, the importance of several other modes has   been
emphasised.  By exploiting techniques of
flavour identification of $b$-jets, thereby reducing the huge QCD
background from light-quark and gluon jets, the modes
 (iv) $q\bar q'\ar WH\ar \ell\nu_\ell b\bar
b$ and (v) $gg,q\bar q\ar t\bar t H\ar b\bar b b\bar b WW
\ar b\bar b b\bar b  \ell\nu_\ell X$, can be used to search 
for the \sm\ Higgs \cite{Tevadetect1,Tevadetect2}. 
Another potentially important channel, particularly
for the mass range $2 M_W \Ord M_H  \Ord 2 M_Z$,  is
(vi) $ H\ar W^{(*)}W^{(*)}\ar \ell^+\nu_\ell \ell'^-\bar\nu_{\ell'}$
 \cite{dittdrei}. Here the lack of a measurable narrow resonant
peak is compensated by a relatively large branching ratio,
since for this mass range $H\ar WW$ is the dominant decay mode.

In Figs.~\neventgg--\neventlnln\ we show the product of the cross sections
and  branching ratios
for the above channels, again at  10 and 14~TeV, for the Higgs mass ranges
where they give sizeable event rates (in Figs.~\neventwh(b) and
\neventwhbb(b) $m_t$ is set equal to 175 GeV).
As already noticed in Ref.~\cite{oldpaper}, the combination of a
rising $H\ar \gamma\gamma$ branching ratio with a falling cross
section yields, for  cases (i) and (ii) above, 
 a remarkably constant signal for $M_H\Ord 140$~GeV.

Finally, we should also mention 
the channel: (vii) $H\ar Z Z\ar \ell\bar \ell\nu\bar\nu$,
with $\ell=e,\mu$ and $e,\mu,\tau$-neutrinos, since it may offer 
additional chances for Higgs detection
in the very heavy mass range \cite{ATLAS,CMS}.\footnote{When the
Higgs resonance becomes very broad, for large $M(H)$, 
 the $p_T(Z\to \ell\bar \ell)$
spectrum may give a cleaner signal.}
The corresponding event rates can be obtained simply by multiplying the
numbers of Fig.~13a--b by six.

\section{Conclusions}
\label{sec:conc}

In this paper we have computed all the decay modes
and the most important production mechanisms of the \sm\ Higgs at the
LHC, by using the most recent sets of parton distributions and by
including in our computations all the available next-to-leading
order corrections.
Cross sections have been presented for two  values of
 the LHC collider energy, $\sqrt s_{pp}=10$
and 14~TeV.
As the most promising
signatures which should allow for Higgs detection at the LHC are
\begin{itemize}
\item $gg\ar H\ar \gamma\gamma$,
\item $q\bar q'\ar WH\ar \ell\nu_\ell\gamma\gamma$
and $gg,q\bar q\ar t\bar t H\ar \ell\nu_\ell\gamma\gamma X$,
\item $q\bar q'\ar WH\ar \ell\nu_\ell b\bar
b$ and $gg,q\bar q\ar t\bar t H\ar b\bar b b\bar b WW
\ar b\bar b b\bar b  \ell\nu_\ell X$,
\item $gg\ar H\ar
Z^{(*)}Z^{(*)}\ar \ell^+\ell^-\ell'^{+}\ell'^{-}$, where
$\ell,\ell'=e$ or $\mu$,
\item $gg\ar H\ar
W^{(*)}W^{(*)}\ar \ell^+\nu_\ell \ell'^{-}\bar\nu_{\ell'}$,
where $\ell,\ell'=e$ or $\mu$,\footnote{In the analysis of
Ref.~\cite{dittdrei} the additional decay channels $W\ar \tau \nu
\ar (e,\mu)+\nu$'s were also included, yielding a slightly larger
signal event rate.}
\item $gg\ar H\ar ZZ\ar \ell^+\ell^-\nu_{\ell'}\bar\nu_{\ell'}$,
where $\ell=e$ or $\mu$ and $\ell'=e,\mu$ or $\tau$,
\end{itemize}
we have presented
updated numbers for the corresponding event rates.
The theoretical uncertainty of the results, which mainly arises from
\begin{itemize}
\item the lack of precise knowledge  of the gluon
distribution at small $x$ (which is particularly important
for the intermediate mass Higgs case, the most difficult to recognise
at the LHC),
\item the uncertainty in the value of $\alpha_s(M_Z)$,
\item the effect of unknown  higher-order perturbative QCD corrections
as well as the scale dependence of those already computed,
\end{itemize}
has been investigated and estimated,
by adopting different sets of recent NLO parton
distributions and comparing their results, and by studying the
(renormalisation and factorisation) scale dependence, at NLO, of the most
important Higgs production channel (via gluon-gluon fusion) at the LHC.
We estimate the current theoretical errors  to 
be $\approx \pm 20\%$ (for the uncertainty
due to parton distributions and $\alpha_s$) and 
$\approx\pm 30\%$ (for the error due to the scale dependence, see also
Ref.~\cite{sdgz}),
the latter for the gluon-gluon fusion process.

In summary, the values presented here for branching ratios, cross sections and
event rates correspond to the state-of-the-art in our current 
knowledge of the input quantities and higher-order corrections, 
 and should be a useful reference for the normalisations
used in the various experimental simulations.

\section*{Acknowledgements}
We thank Michael Spira for useful comments and suggestions.
SM and WJS are grateful to the UK PPARC for support.
This work was supported in part by the EU Programme
``Human Capital and Mobility'', Network ``Physics at High Energy
Colliders'', contract CHRX-CT93-0537 (DG 12 COMA).

\vskip0.5cm\noindent
{{\sl Note added}}. {After submitting this manuscript to the journal
we received a private communication from Michael Spira in which he confirmed 
the correctness of our speculation about the discrepancies between the 
approximate analytical result and the numerical complete one in the scale 
dependence of the $gg\ar H$ cross sections at NLO \cite{higlu}.}

\goodbreak

\section*{Figure captions}

\begin{itemize}

\item[{[\brlo]}] Branching ratios of the \sm\ Higgs boson in the mass
range 50~GeV $< M_H <$~200 GeV, for the decay modes: a) $b\bar b$,
$c\bar c$, $\tau^+\tau^-$, $\mu^+\mu^-$ and $gg$; b) $WW$,
$ZZ$, $\gamma\gamma$ and $Z\gamma$.

\item[{[\brhi]}] Branching ratios of the \sm\ Higgs boson in the mass
range 200~GeV $< M_H <$ 1~TeV, for the decay modes: $WW$,
$ZZ$ and $t\bar t$ ($m_t=175$ GeV).

\item[{[\gammatot]}] Total width of the \sm\ Higgs as a function of the mass
in the range 50~GeV $< M_H <$ 1~TeV  ($m_t=175$ GeV).

\item[{[\feyndiags]}] Representative Feynman diagrams describing the main
mechanisms of Higgs production at the LHC: (a) gluon-gluon fusion; (b)
$WW$- or $ZZ$-fusion; (c) associated production with $W$ or $Z$;
and (d) associated production with top quark pairs. 

\item[{[\sigtot]}] Total cross sections for $H$ production at the  LHC as a
function of the Higgs mass $M_H$, as given by the four production
mechanisms illustrated in Fig.~4, at $\sqrt s_{pp}=10$~TeV (a)
and $\sqrt s_{pp}=14$~TeV (b), and their ratios 
$\sigma(10~{\mathrm {TeV}})/\sigma(14~{\mathrm {TeV}})$ (c). 
Here,  $m_t=175$~GeV.

\item[{[\sigtop]}] The ratios $R_t$ of the total
cross sections for $H$ production, via $gg\ar H$ (continuous curves) and 
$gg, q\bar q\ar t\bar t H$ (dashed curves), as a
function of the Higgs mass $M_H$, at $\sqrt s_{pp}=10$ and $14$~TeV,
for  $m_t=165$ and 175~GeV, and for 
$m_t=185$ and 175~GeV.

\item[{[\sigpartons]}] Ratios (with respect to MRS(A)) of \sm\ Higgs 
production cross sections from gluon-gluon fusion
calculated using eleven  dif\-fe\-re\-nt se\-ts of par\-ton distributions:
 MRS(A,\- A$'$,\- G,\- R1,\- R2), 
 CTEQ(2M,\- 2MS,\- 2MF,\- 2ML,\- 3M) and GRVHO94.
Note that  $m_t=175$~GeV.

\item[{[\gluons]}] Behaviour of the  gluon distributions $xg(x,Q^2)$
of  the  various parton sets used in Fig.~{\sigpartons}
 as a function of $x$, with 
$Q^2=x^2 s_{pp}$ and  $\sqrt s_{pp}=14$~TeV. The ordering
of the labels in the legend corresponds to  increasing
values of $xg$ at the minimum value of $x$.

\item[{[\scaledep]}] Higgs production cross section for $\sqrt{s} = 14$~TeV,
$M_H = 100$~GeV from the gluon-gluon fusion process, as a function
of the (equal) renormalisation and factorisation scales $\mu$, at LO
and NLO order. The GRV94 sets have been used here.
Note that  $m_t=175$ GeV.

\item[{[\neventgg]}] Total cross sections for \sm\ Higgs production 
times the branching
ratio for the decay mode $H\ar \gamma\gamma$, as a function
of the Higgs mass $M_H$ in the low mass range, 
at $\sqrt s_{pp}=10$ TeV and $\sqrt s_{pp}=14$ TeV.

\item[{[\neventwh]}] Total cross sections for $WH$ (continuous curves) and
$ZH$ production (dashed curves) (a) and for
$t\bar tH$ (continuous curves) (b)
times the branching ratios of the decay
modes $H\ar \gamma\gamma$, $W\ar\ell\nu_\ell$, $Z\ar
\ell^+\ell^-$ and $t\bar t\ar \ell\nu_\ell X$ ($\ell=e,\mu$), as a function
of the Higgs mass $M_H$ in the low mass range, 
at $\sqrt s_{pp}=10$ TeV and $\sqrt s_{pp}=14$ TeV.
Note that  $m_t=175$ GeV.

\item[{[\neventwhbb]}] Total cross sections for $WH$ and
$ZH$ production (a) and for
$t\bar tH$ (b)
times the branching ratios of the decay
modes $H\ar b\bar b$, $W\ar\ell\nu_\ell$, $Z\ar
\ell^+\ell^-$ and $t\bar t\ar \ell\nu_\ell X$ ($\ell=e,\mu$), as a function
of the Higgs mass $M_H$ in the low mass range, 
at $\sqrt s_{pp}=10$ TeV and $\sqrt s_{pp}=14$ TeV.
Note that  $m_t=175$ GeV.

\item[{[\neventllll]}] Total cross sections 
for \sm\ Higgs production times the branching
ratio for the decay mode $H\ar Z^{(*)} Z^{(*)} \ar
\ell^+\ell^-\ell^+\ell^-$ ($\ell=e,\mu$),  as a function
of the Higgs mass in the ranges (a)  50~GeV $ \leq M_H \leq  300$~GeV and (b)
 100~GeV $\leq M_H \leq  1$~TeV,
at $\sqrt s_{pp}=10$ TeV and $\sqrt s_{pp}=14$~TeV.
Note that  $m_t=175$ GeV.

\item[{[\neventlnln]}] Total cross sections for \sm\
Higgs production times the branching
ratio for the decay mode $H\ar W^{(*)} W^{(*)} \ar
\ell^+\nu_\ell\ell'^-\bar\nu_{\ell'}$ ($\ell,\ell'=e,\mu$),  as a function
of the Higgs mass in the range
  $0 \leq M_H \leq  300$~GeV,
at $\sqrt s_{pp}=10$ TeV and $\sqrt s_{pp}=14$~TeV.
Note that  $m_t=175$ GeV.

\end{itemize}
\vfill
\newpage

\begin{figure}[p]
~\epsfig{file=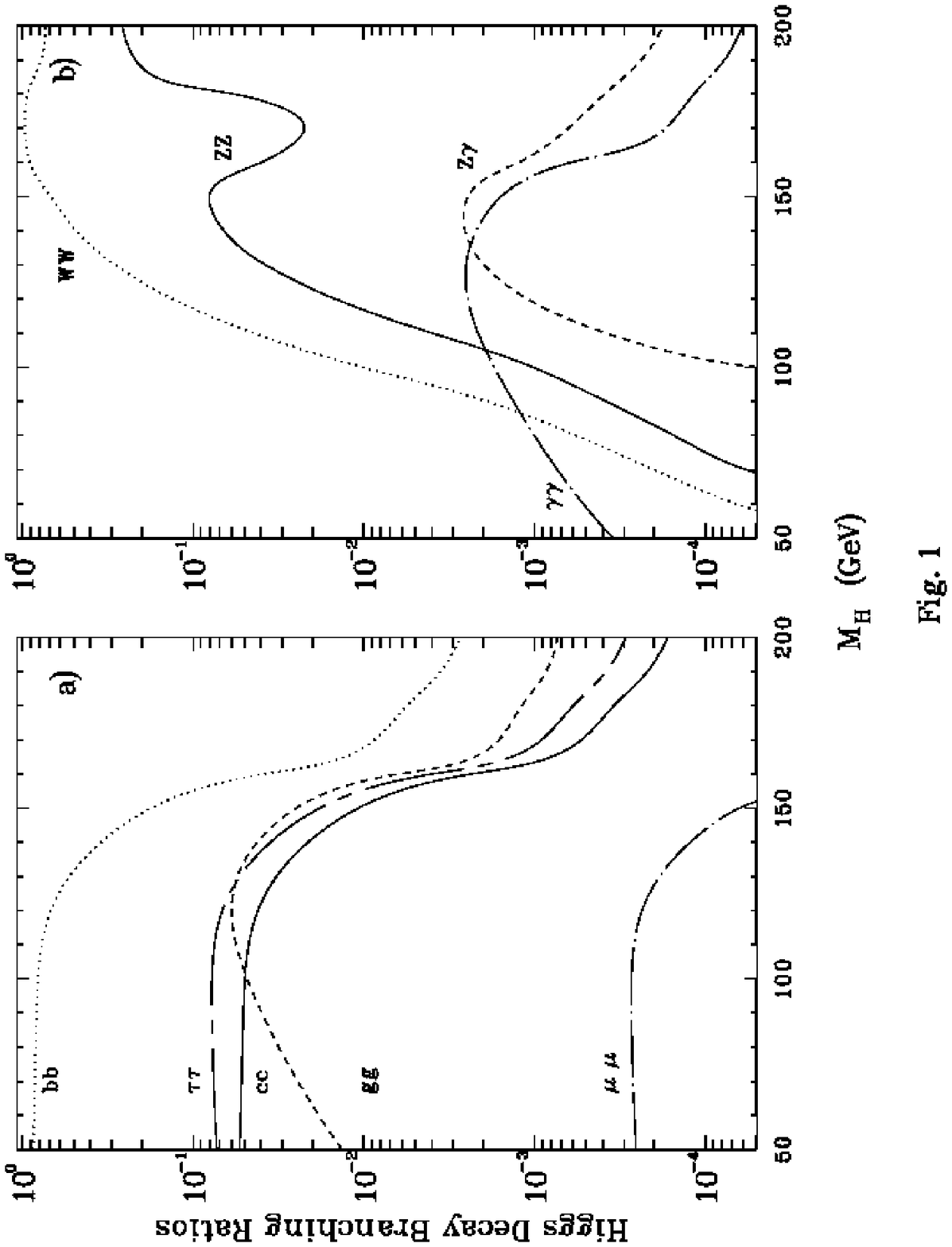,height=22cm}
\end{figure}
\stepcounter{figure}
\vfill
\clearpage

\begin{figure}[p]
~\epsfig{file=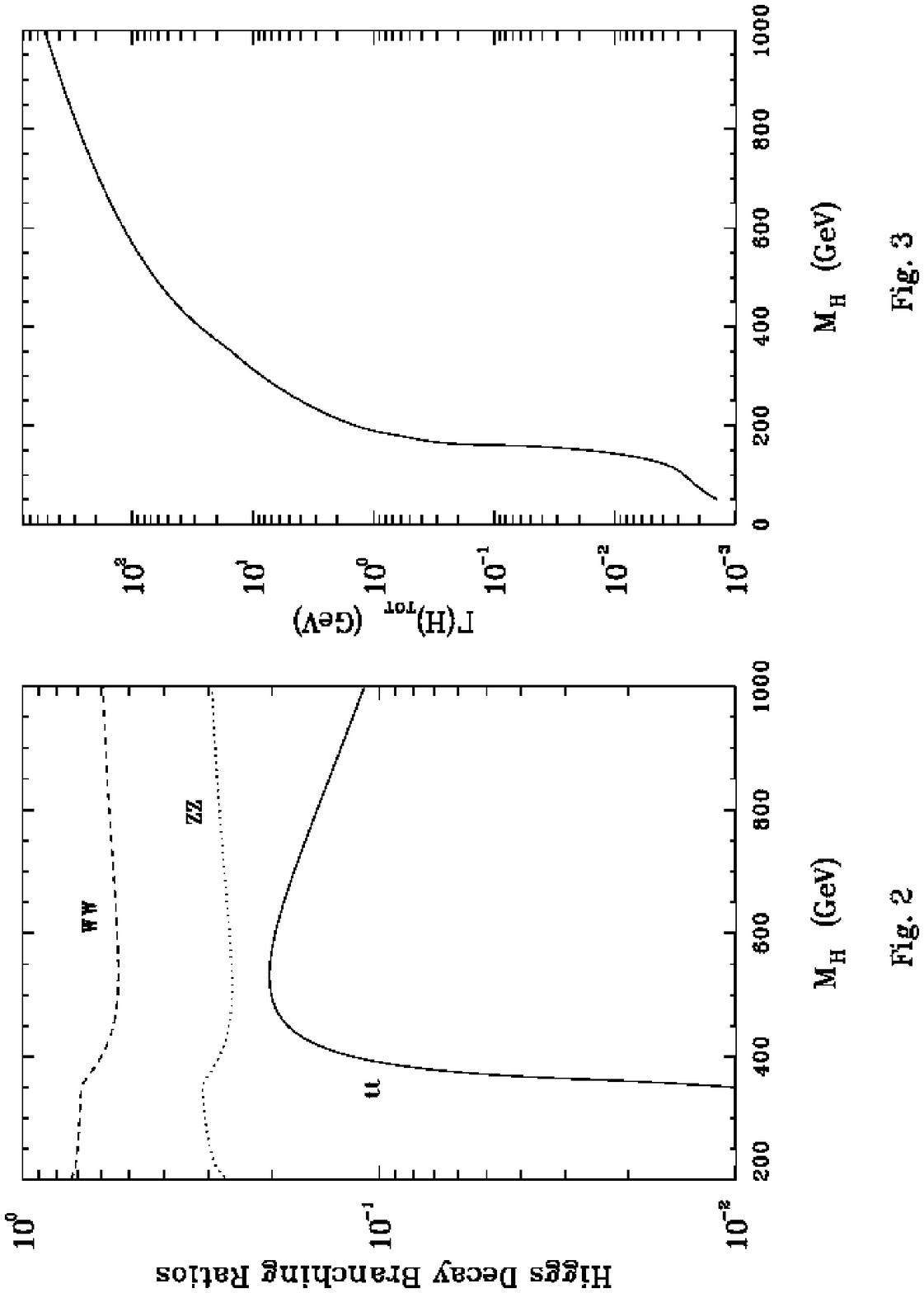,height=22cm}
\end{figure}
\stepcounter{figure}
\vfill
\clearpage

\begin{figure}[p]
~\epsfig{file=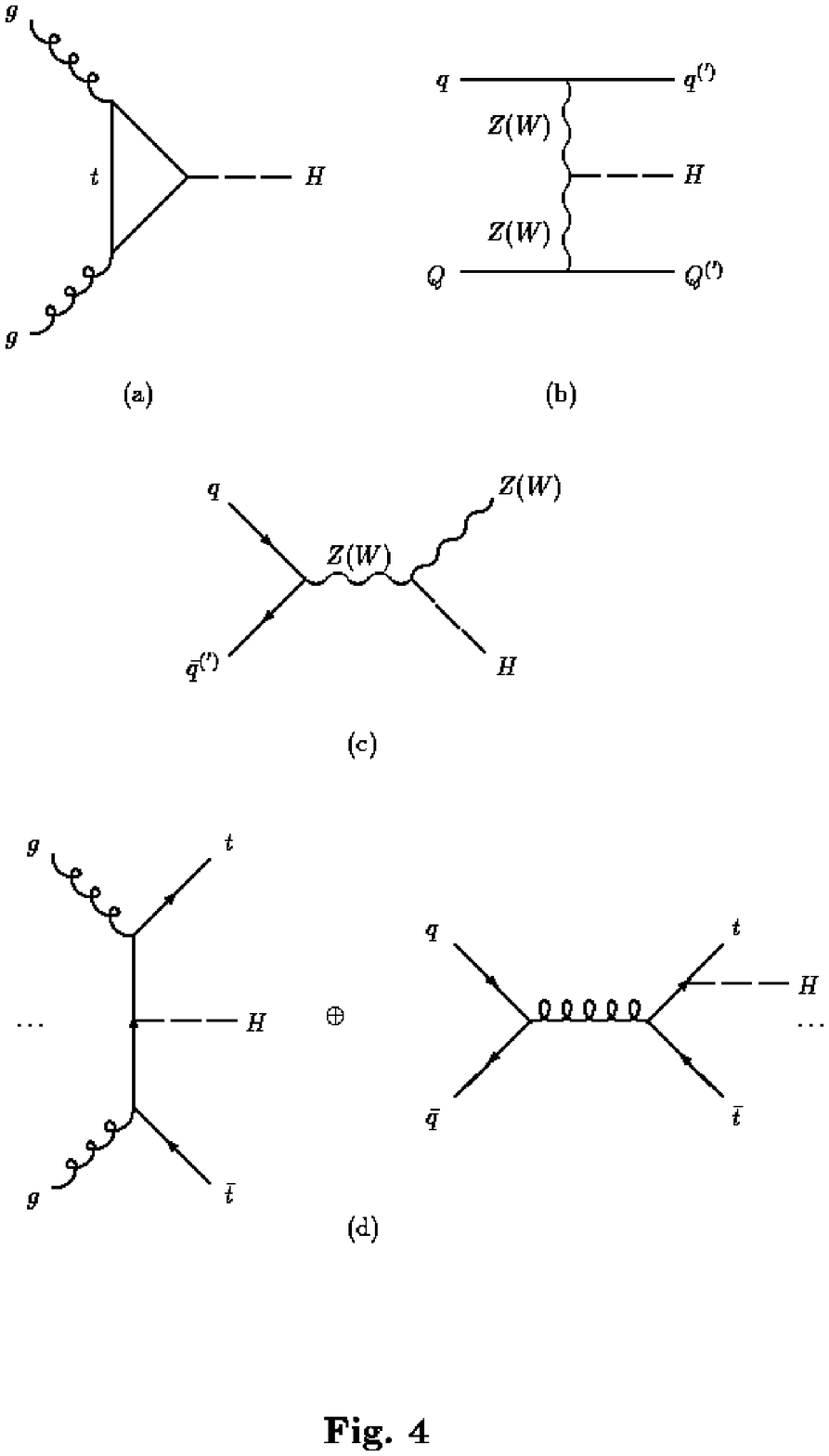,height=22cm}
\end{figure}
\stepcounter{figure}
\vfill
\clearpage

\begin{figure}[p]
~\epsfig{file=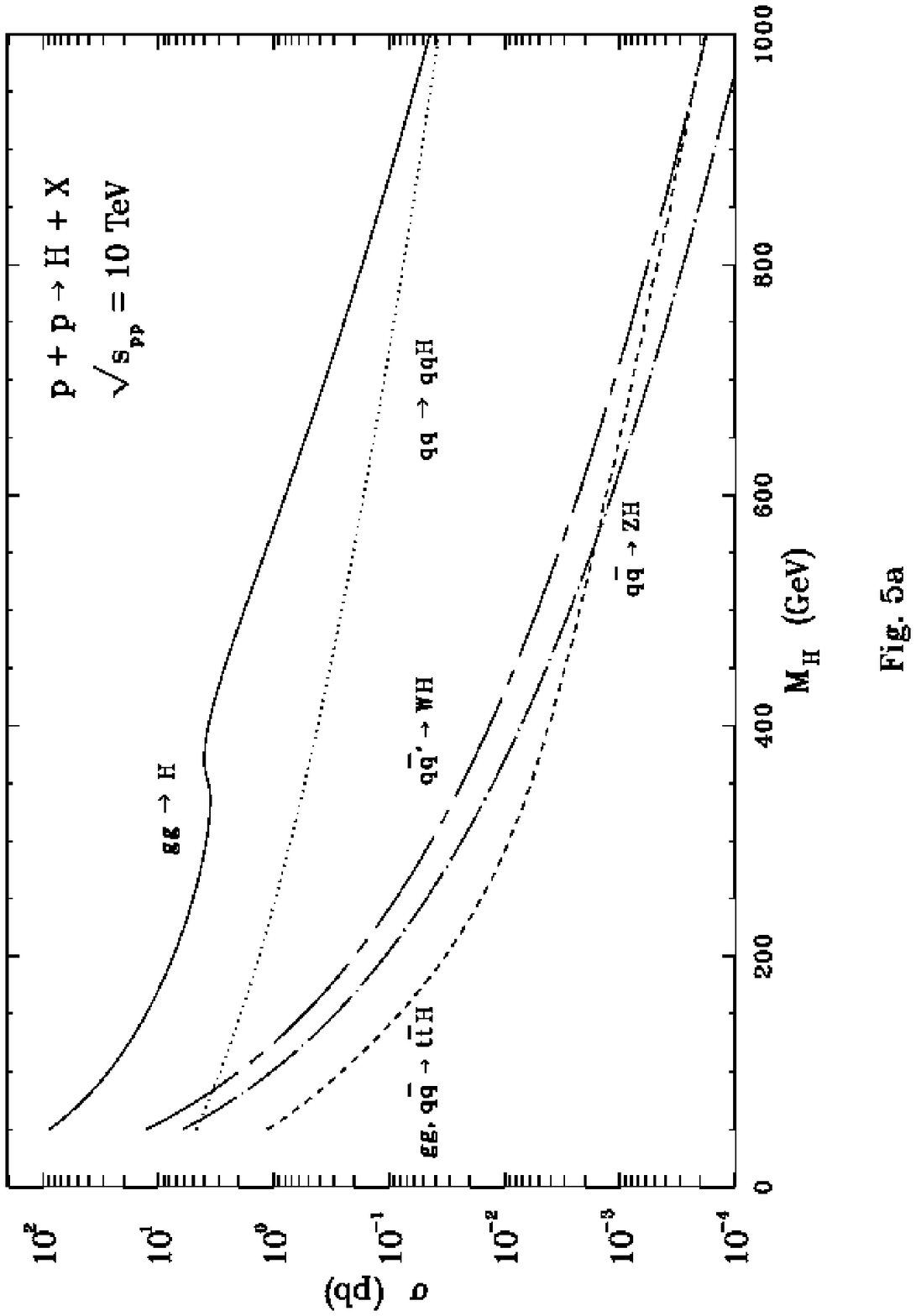,height=22cm}
\end{figure}
\stepcounter{figure}
\vfill
\clearpage

\begin{figure}[p]
~\epsfig{file=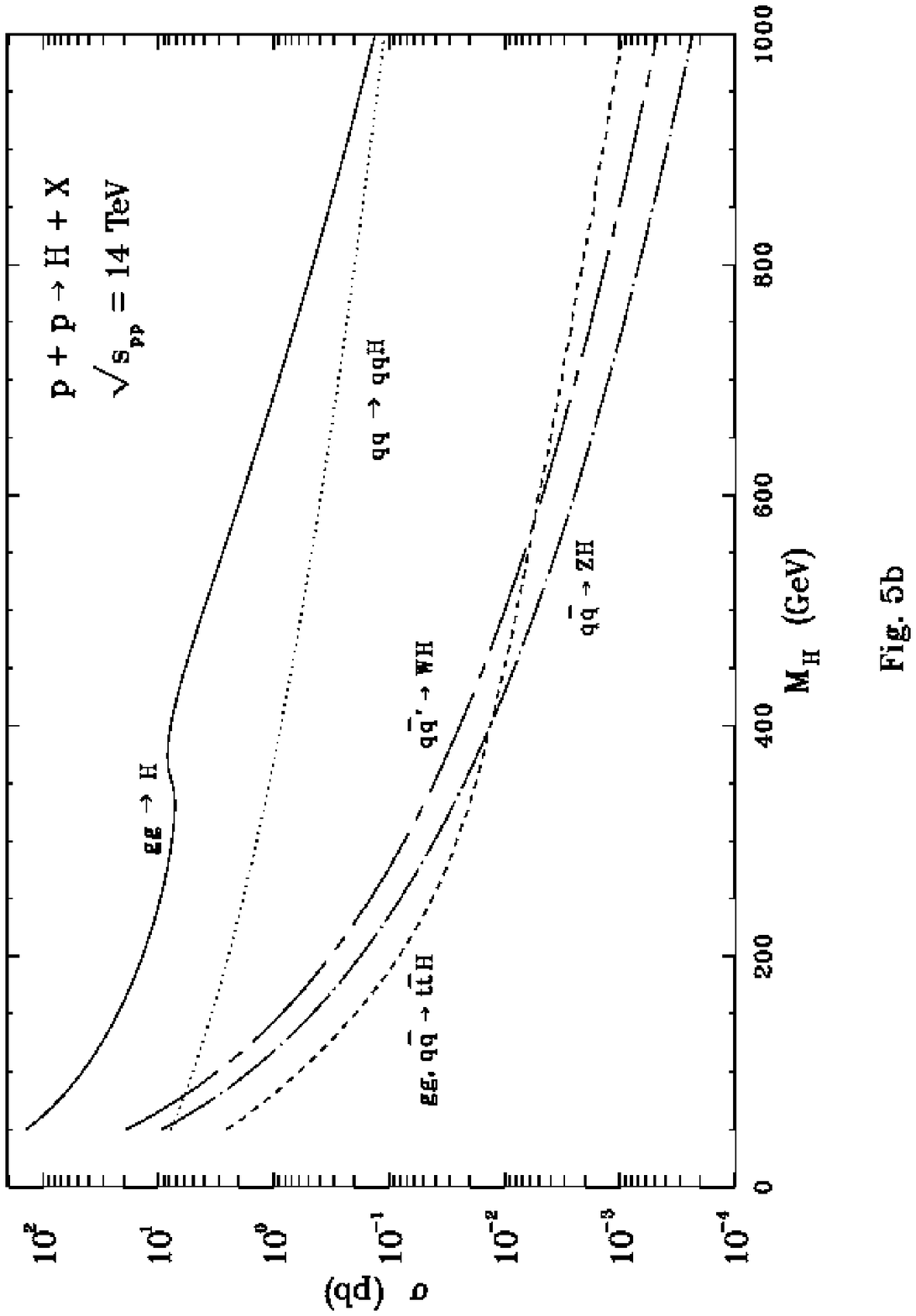,height=22cm}
\end{figure}
\stepcounter{figure}
\vfill
\clearpage

\begin{figure}[p]
~\epsfig{file=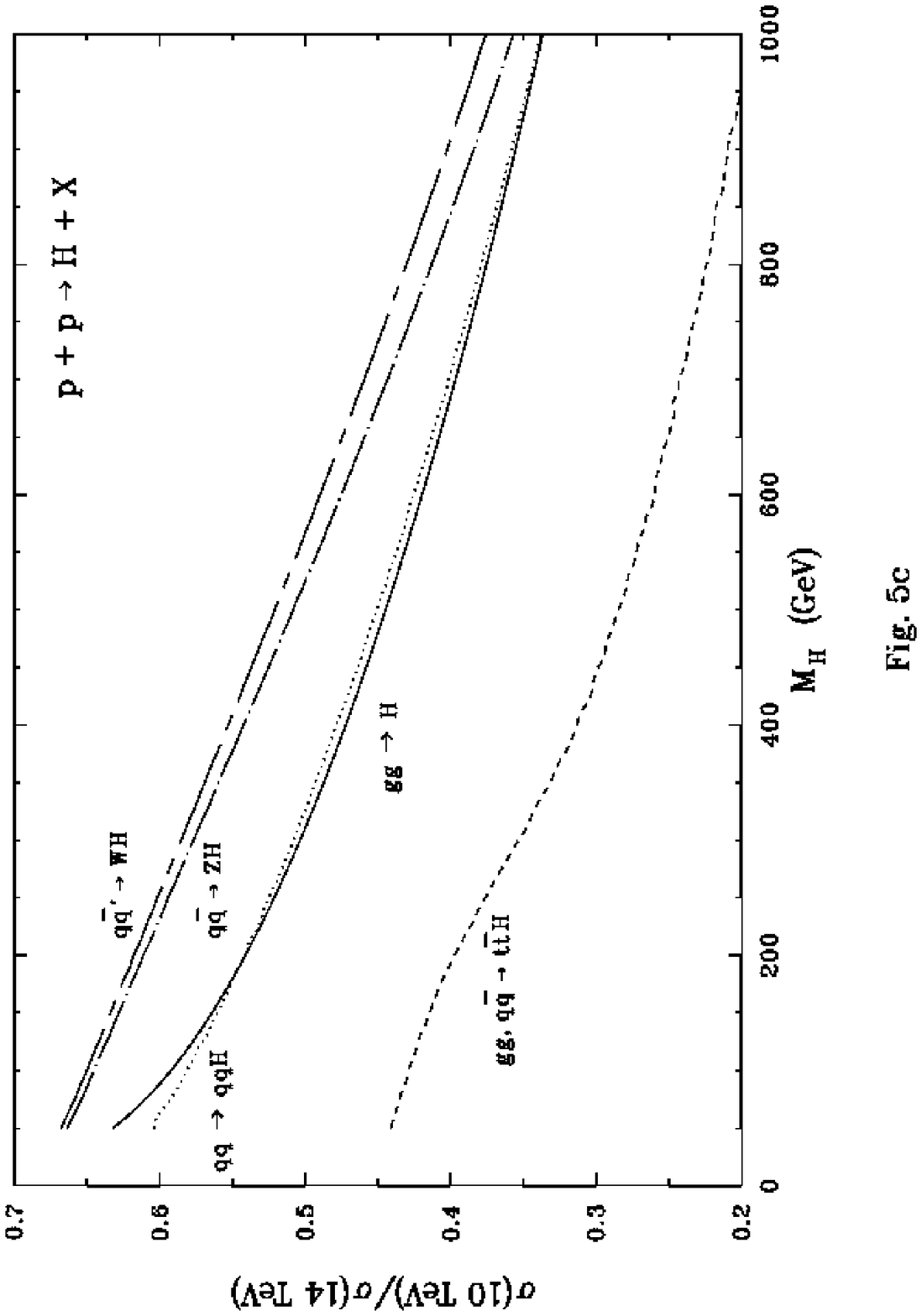,height=22cm}
\end{figure}
\stepcounter{figure}
\vfill
\clearpage

\begin{figure}[p]
~\epsfig{file=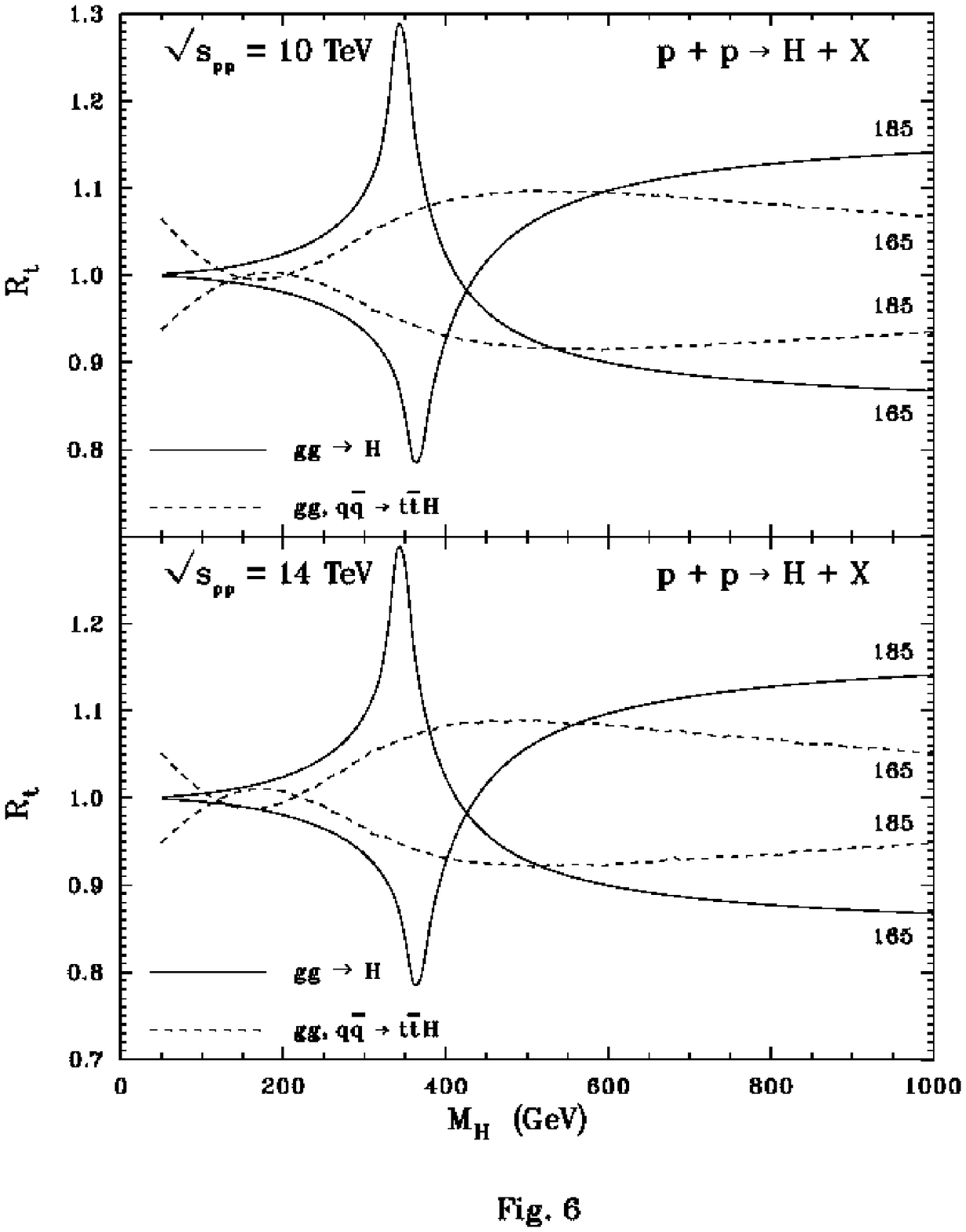,height=22cm}
\end{figure}
\stepcounter{figure}
\vfill
\clearpage

\begin{figure}[p]
~\epsfig{file=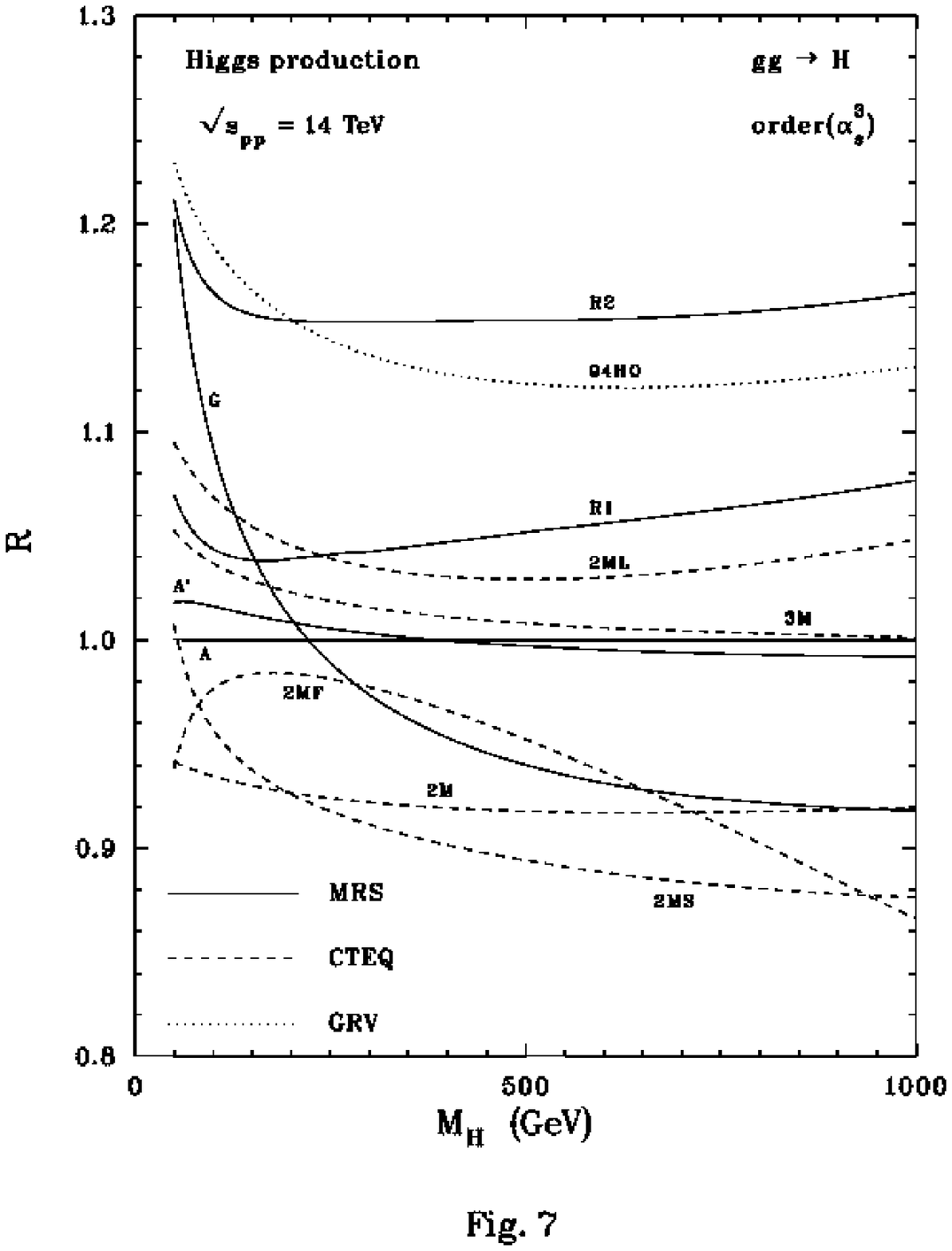,height=22cm}
\end{figure}
\stepcounter{figure}
\vfill
\clearpage

\begin{figure}[p]
~\epsfig{file=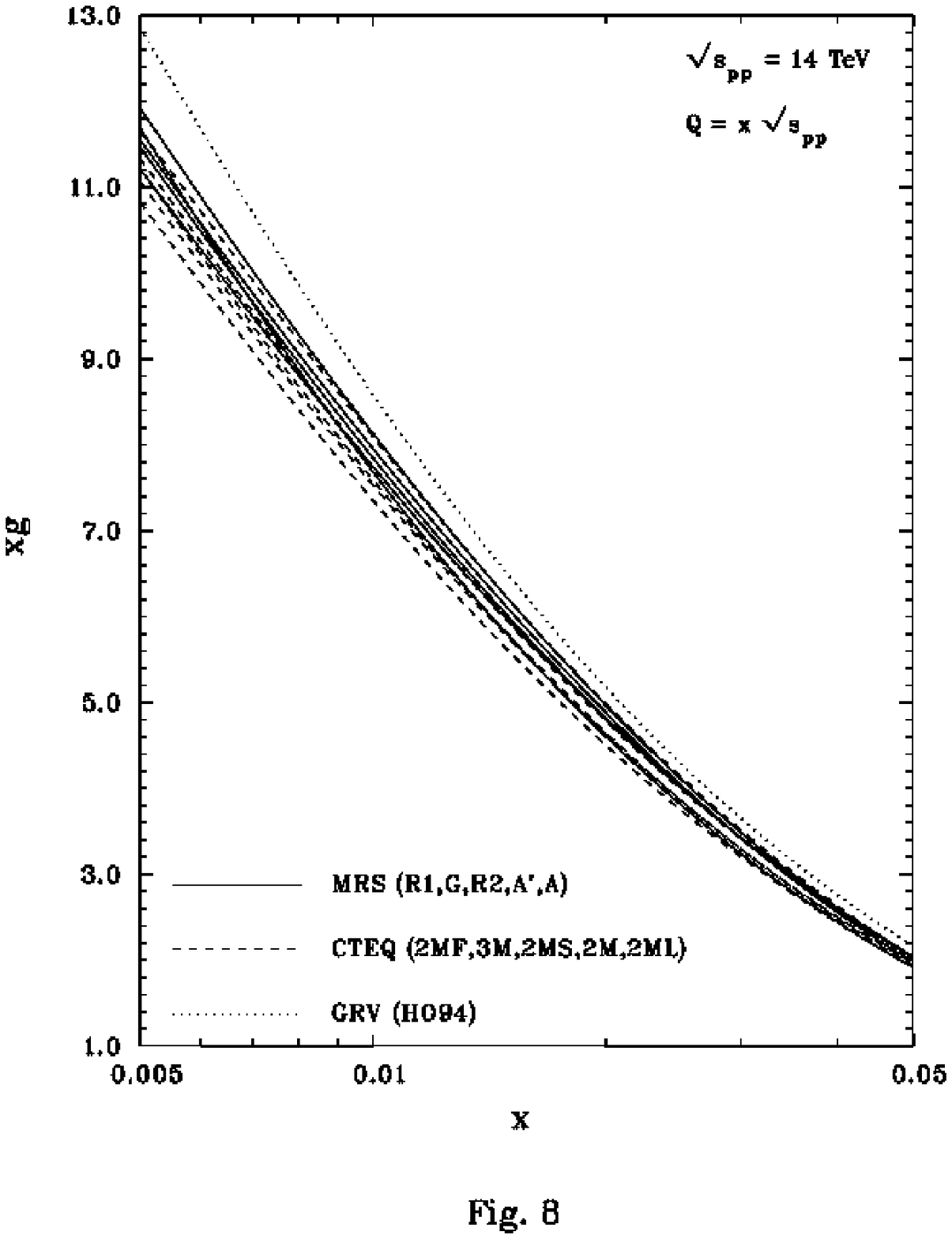,height=22cm}
\end{figure}
\stepcounter{figure}
\vfill
\clearpage

\begin{figure}[p]
~\epsfig{file=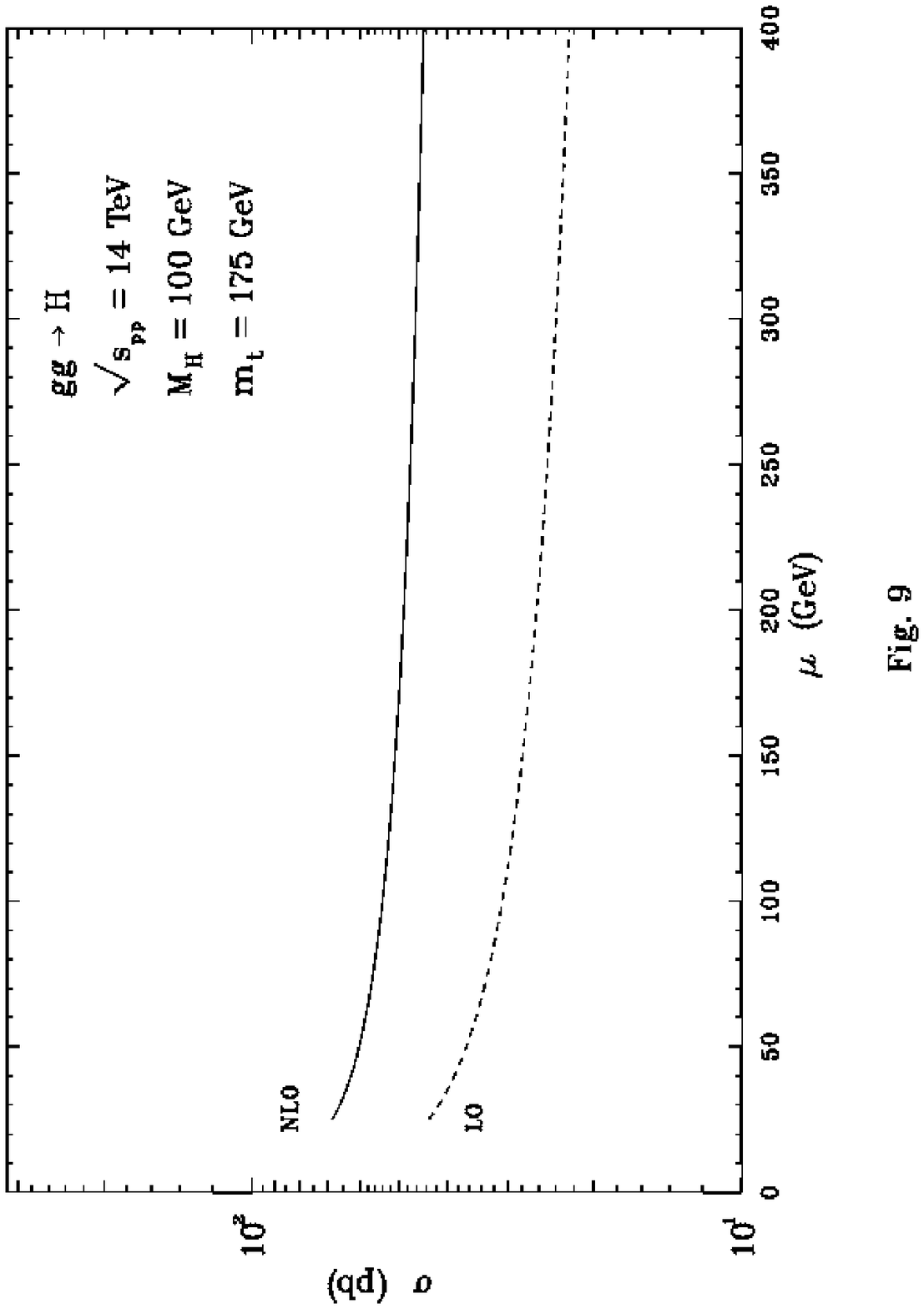,height=22cm}
\end{figure}
\stepcounter{figure}
\vfill
\clearpage

\begin{figure}[p]
~\epsfig{file=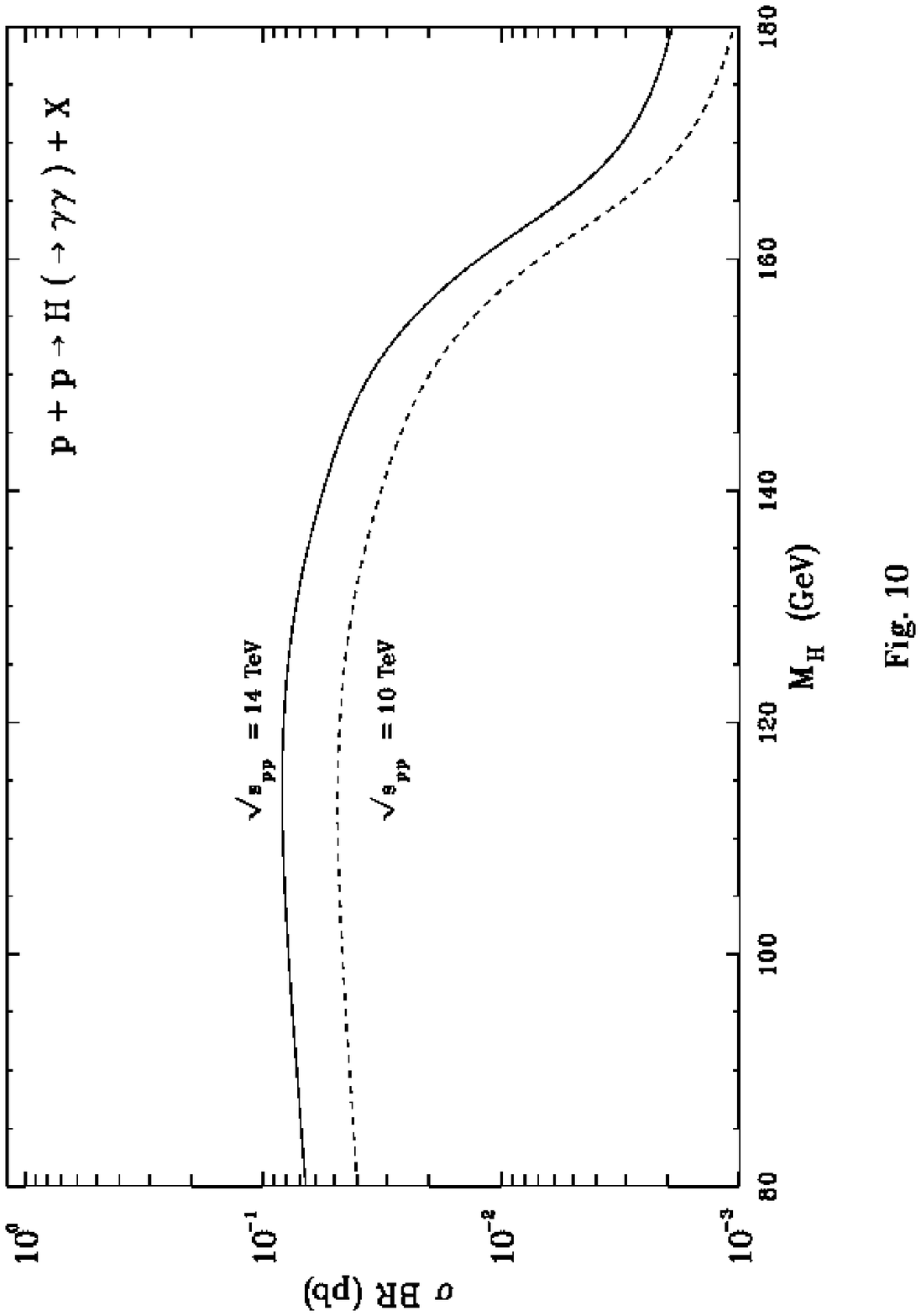,height=22cm}
\end{figure}
\stepcounter{figure}
\vfill
\clearpage

\begin{figure}[p]
~\epsfig{file=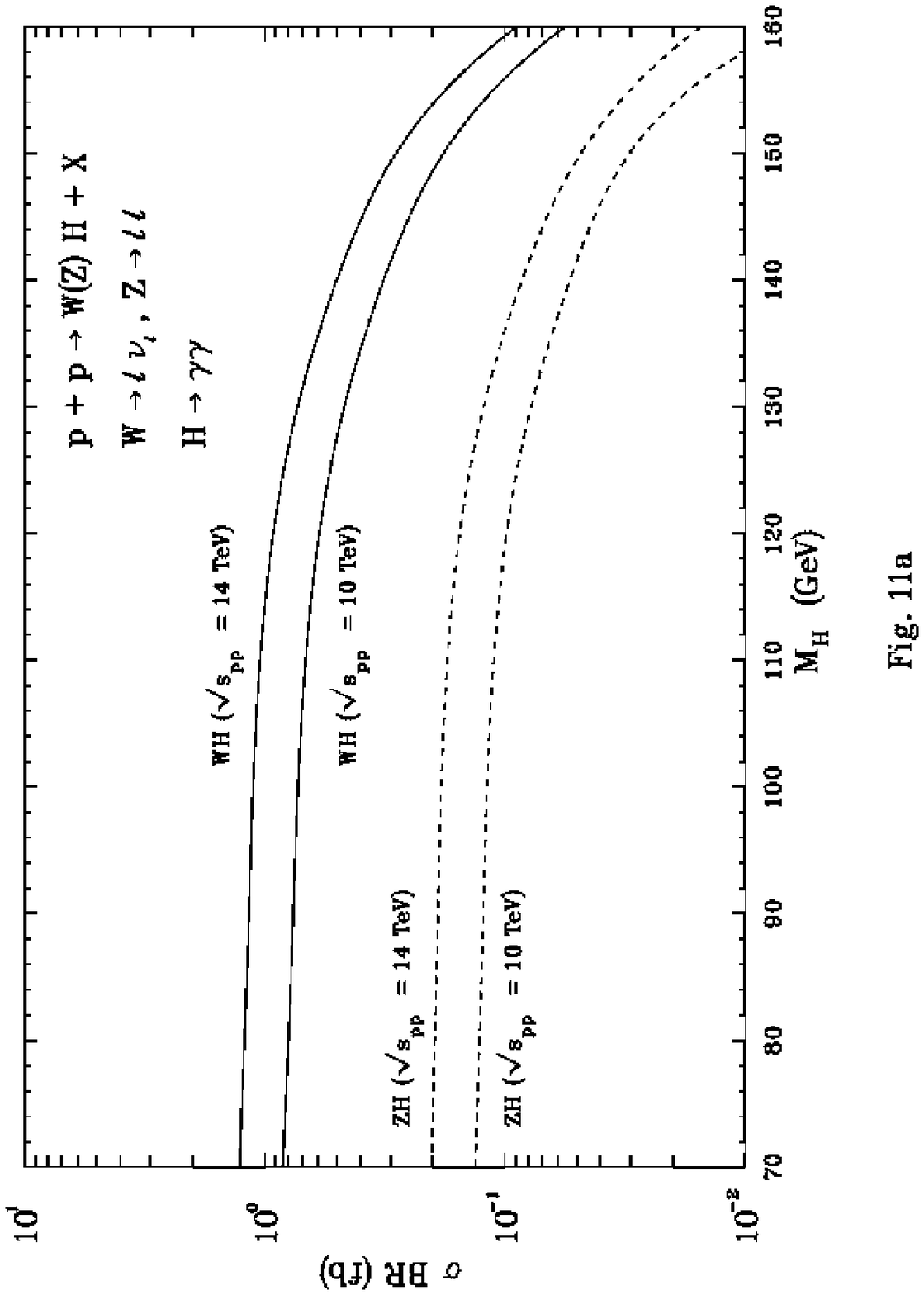,height=22cm}
\end{figure}
\stepcounter{figure}
\vfill
\clearpage

\begin{figure}[p]
~\epsfig{file=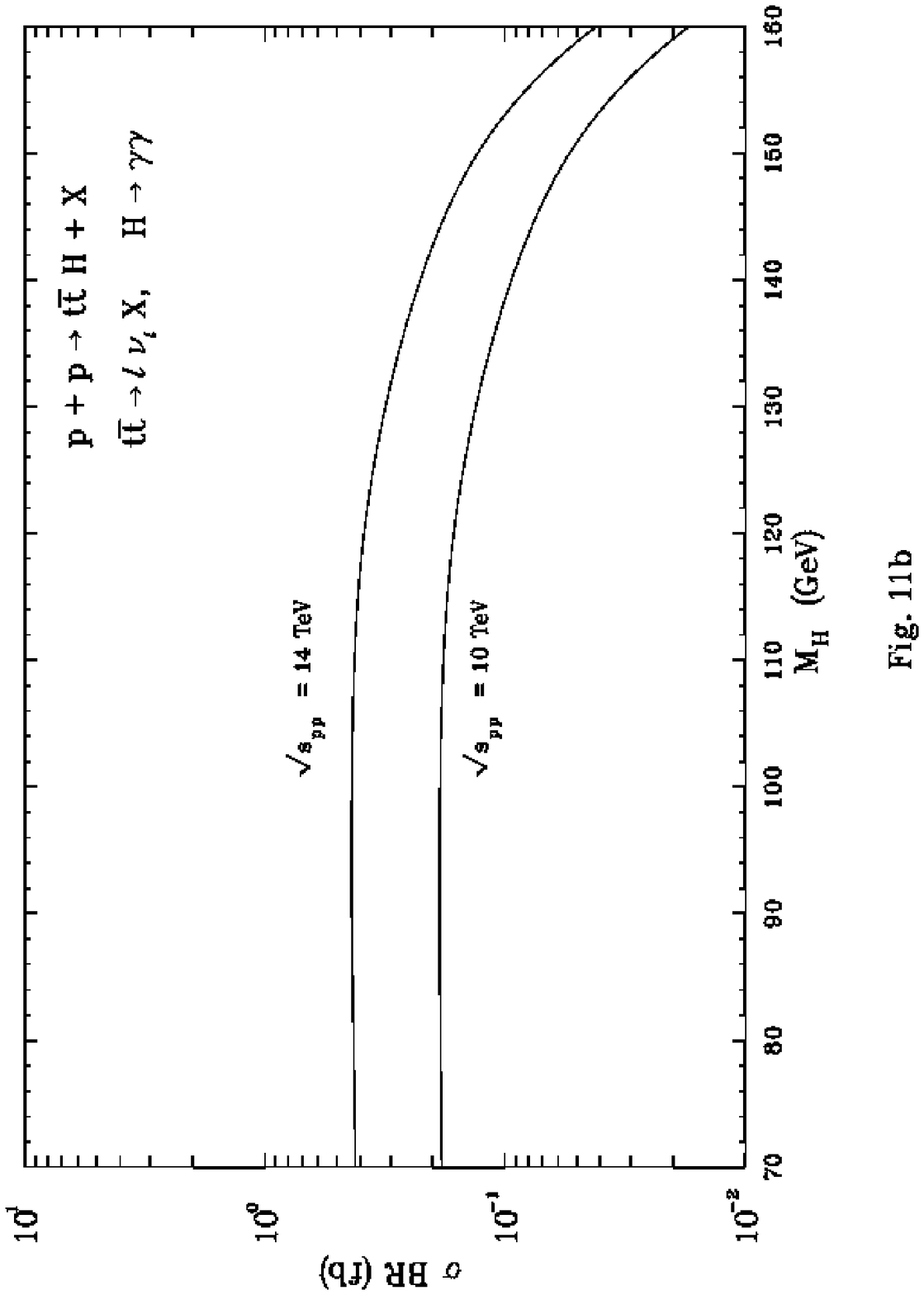,height=22cm}
\end{figure}
\stepcounter{figure}
\vfill
\clearpage

\begin{figure}[p]
~\epsfig{file=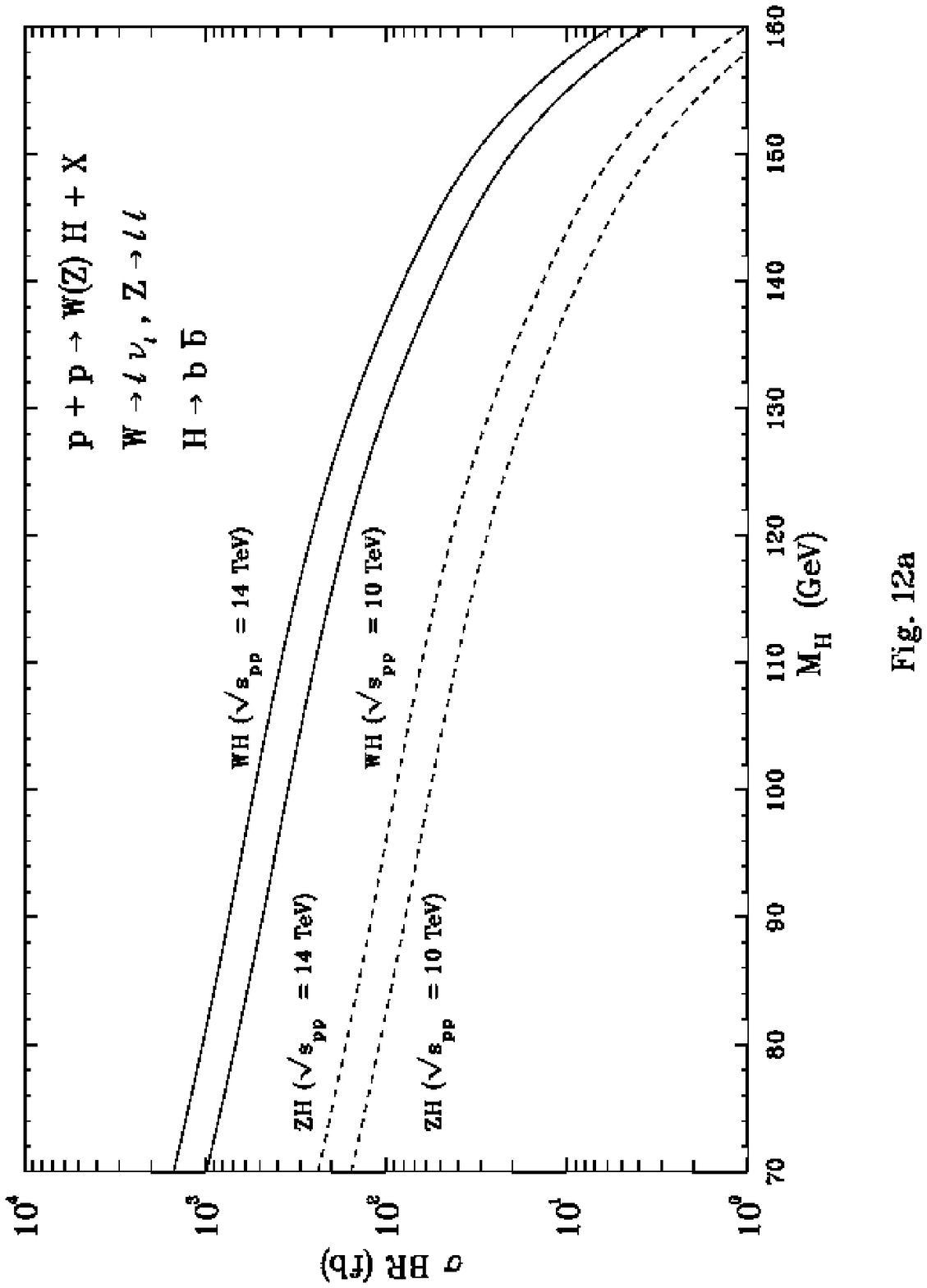,height=22cm}
\end{figure}
\stepcounter{figure}
\vfill
\clearpage

\begin{figure}[p]
~\epsfig{file=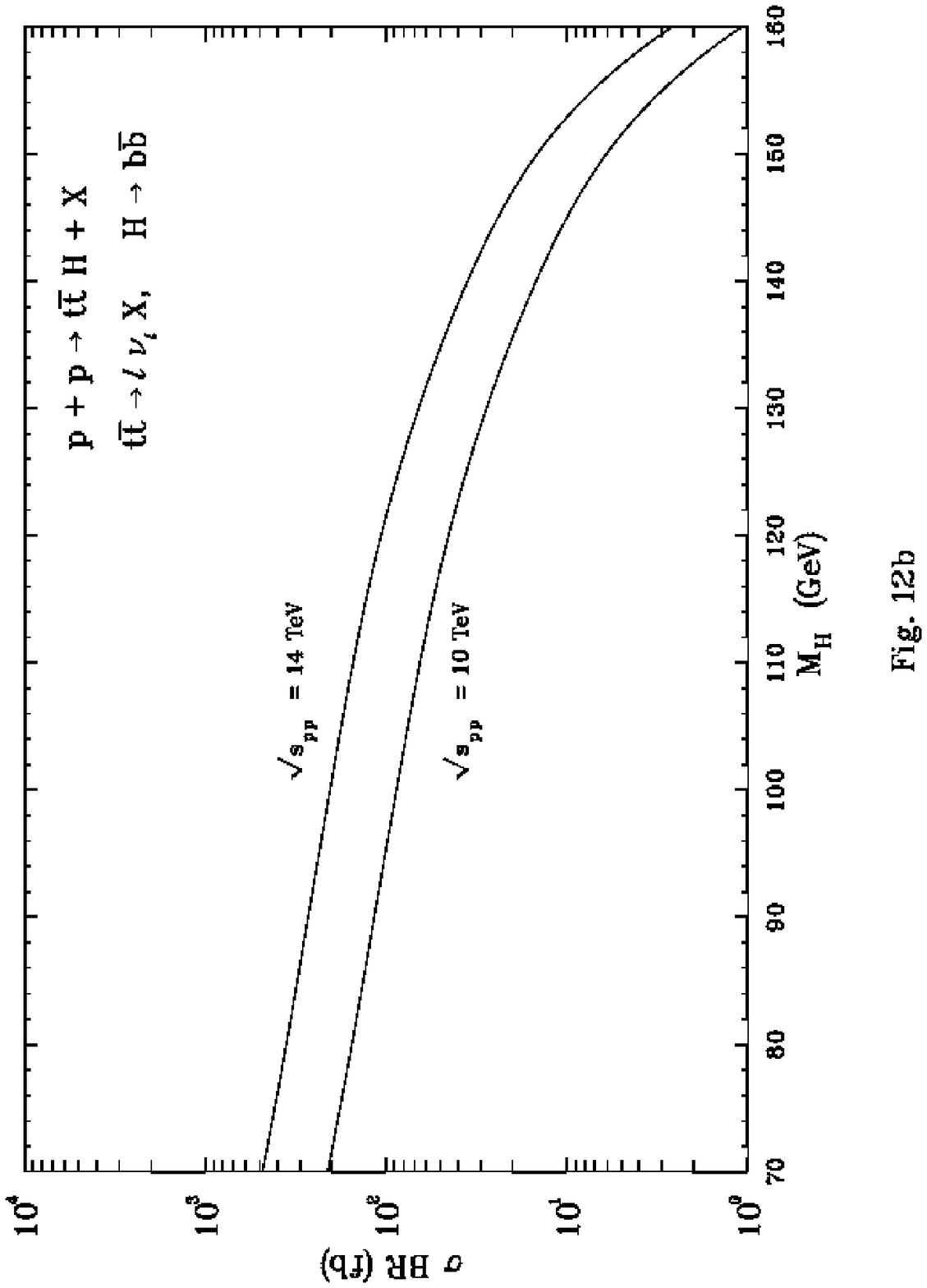,height=22cm}
\end{figure}
\stepcounter{figure}
\vfill
\clearpage

\begin{figure}[p]
~\epsfig{file=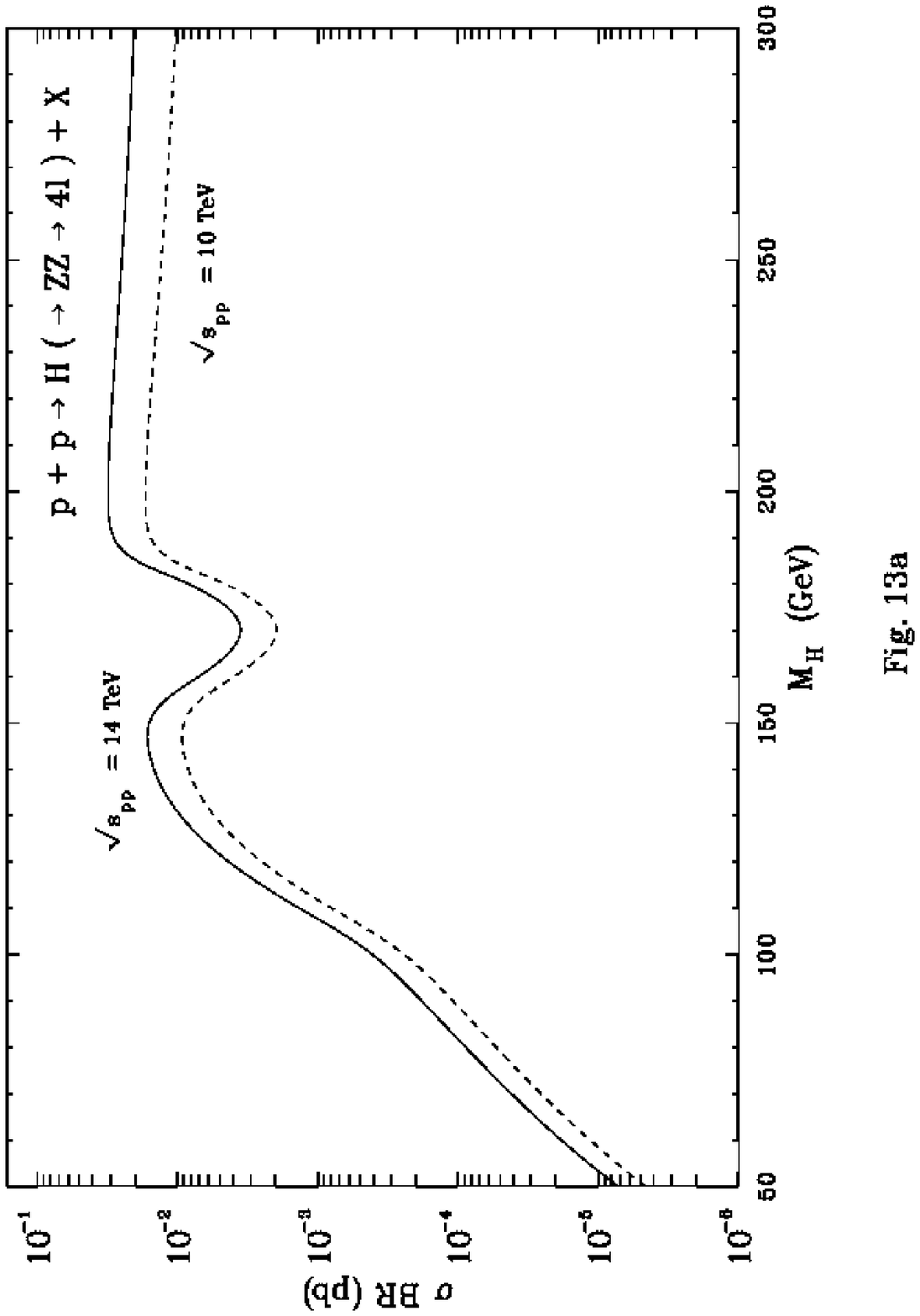,height=22cm}
\end{figure}
\stepcounter{figure}
\vfill
\clearpage

\begin{figure}[p]
~\epsfig{file=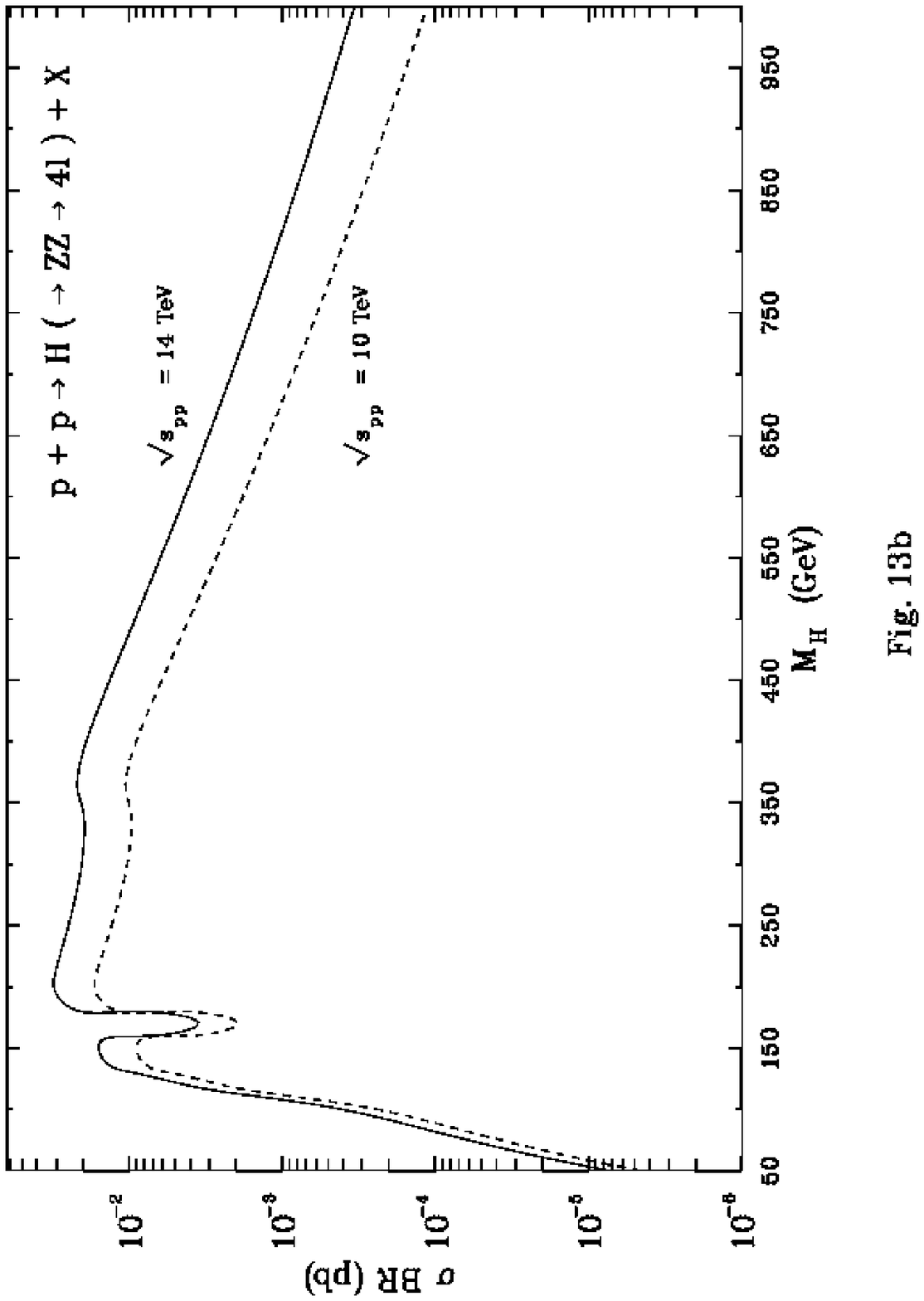,height=22cm}
\end{figure}
\stepcounter{figure}
\vfill

\begin{figure}[p]
~\epsfig{file=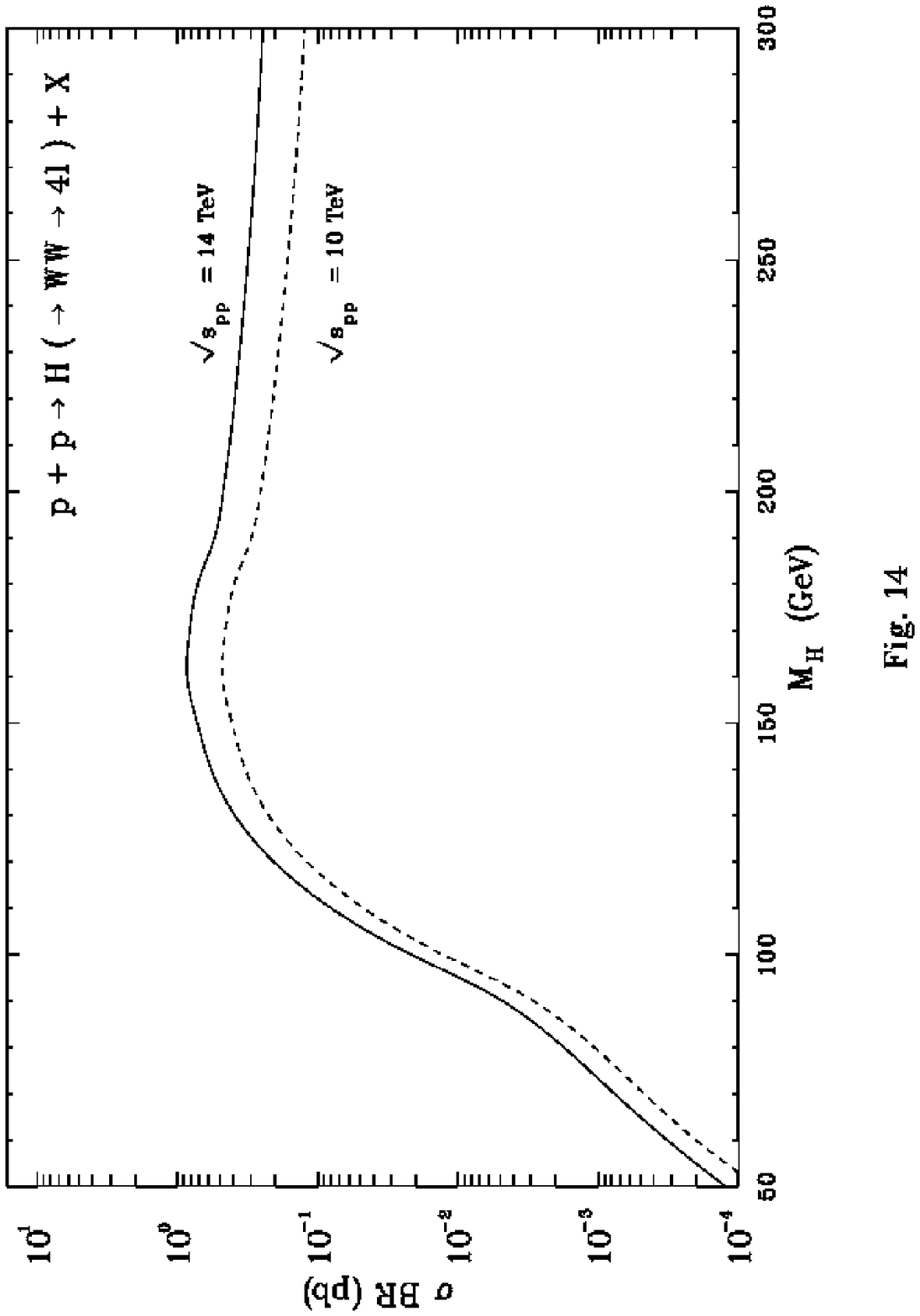,height=22cm}
\end{figure}
\stepcounter{figure}
\vfill

\end{document}